%% file: main.tex
\documentclass[11pt, a4paper]{article}

\usepackage[a4paper, top=2.5cm, bottom=2.5cm, left=2cm, right=2cm]{geometry}
\usepackage[english]{babel}
\usepackage{graphicx}
\usepackage{amsmath}
\usepackage{amssymb}
\usepackage{hyperref}
\usepackage{cite}
\usepackage{lipsum}
\usepackage{subcaption}
\usepackage{tikz}
\usepackage{tikz-feynman}
\usepackage{abstract}
\usepackage{authblk}
\usepackage{booktabs}
\usepackage{tabularx}
\usepackage{makecell}
\usepackage{multirow}
\usepackage{lineno} 
\usepackage[normalem]{ulem}

\babelprovide[import, onchar=ids fonts]{english}
\makeatletter
\@ifundefined{ruledtabular}{%
  \newenvironment{ruledtabular}{}{}%
}{}
\makeatother

\title{
  \textbf{  
    Searching for Lepton Flavor Violating decays of the Higgs Boson into $\mu\tau$, $e\tau$, and $e\mu$ final states at FCC-ee
  }
}

\author[1]{P. Sriling}
\author[1,2]{N. Srimanobhas}
\author[3]{P. Uttayarat}
\author[1]{R. Uttho}
\author[1]{V. Wachirapusitanand}

\affil[1]{Center of Excellence in High Energy Physics, Department of Physics, Faculty of Science, Chulalongkorn University, Bangkok 10330, Thailand}
\affil[2]{CERN, European Organization for Nuclear Research, CH-1211 Geneva 23, Switzerland}
\affil[3]{Theoretical High-Energy Physics and Astrophysics Research Unit, Department of Physics, Srinakharinwirot University, 114 Sukhumvit 23 Road, Wattana, Bangkok 10110, Thailand}
\date{\today}

\input{defineCommand}

\input{defineFigures}

\begin{document}


\maketitle
\begin{abstract}
    \noindent 
    \input{abstract}
    \vspace{1cm} 
\end{abstract}

\input{introduction}


\input{simulation}

\input{event_reconstruction}

\input{event_selections}


\input{results}

\input{model}

\input{summary.tex}

\section*{Acknowledgment}
PU thanks Reinard Primulando for useful comments and discussions.
We have received support from the National Science, Research and Innovation Fund program IND\_FF\_68\_369\_2300\_097, and the Program Management Unit for Human Resources \& Institutional Development, Research and Innovation, grant B39G680009 (Thailand). The authors acknowledge the National Science and Technology Development Agency, National e-Science Infrastructure Consortium, Chulalongkorn University, and the Chulalongkorn Academic Advancement into Its 2nd Century Project, NSRF via the Program Management Unit for Human Resources \& Institutional Development, Research and Innovation (Thailand) for providing computing infrastructure that has contributed to the research results reported within this paper.

\bibliographystyle{ieeetr}  
\bibliography{bibtex}

\end{document}

%% file: defineCommand.tex
\newcommand{\pt}{p_{\mathrm{T}}}
\newcommand{\htomutau}{H \rightarrow \mu\tau}
\newcommand{\htomutaue}{H \rightarrow \mu\tau_e}
\newcommand{\htoetau}{H \rightarrow e\tau}
\newcommand{\htoetaumu}{H \rightarrow e\tau_\mu}
\newcommand{\htomue}{H \rightarrow \mu e}
\newcommand{\met}{\ensuremath{E_{\mathrm{T}}^{\mathrm{miss}}}}
\newcommand{\mycomment}[1]{}

%% file: defineFigures.tex
\NewDocumentCommand{\ZHVBFdiagram}{}{
    \begin{figure}[!ht]
        \centering
        \begin{subfigure}[b]{0.45\textwidth}
            \centering
            \begin{tikzpicture}
              \begin{feynman}
                \vertex (a); 
                \vertex [right=1.5cm of a] (b); 
                
                \vertex [above left=of a] (i1) {$e^+$};
                \vertex [below left=of a] (i2) {$e^-$};
                
                \vertex [above right=of b] (f1) {$h$};
                \vertex [below right=of b] (f2) {$Z$};
            
                \diagram* {
                  (i1) -- [anti fermion] (a),
                  (i2) -- [fermion] (a),
                  (a) -- [boson, edge label=$Z^*$] (b),
                  (b) -- [scalar] (f1),
                  (b) -- [boson] (f2),
                };
              \end{feynman}
            \end{tikzpicture}
            \caption{Higgs-strahlung ($e^+e^- \to Zh$).}
            \label{fig:ZH_sub}
        \end{subfigure}
        \hfill
        \begin{subfigure}[b]{0.45\textwidth}
            \centering
            \begin{tikzpicture}
              \begin{feynman}
                \vertex (vH); 
                \vertex [right=1.2cm of vH] (fH) {$h$};
                \vertex [above=1.2cm of vH] (v1); 
                \vertex [below=1.2cm of vH] (v2); 
                
                \vertex [above left=0.6cm and 1.5cm of v1] (i1) {$e^+$};
                \vertex [above right=0.6cm and 1.5cm of v1] (f1) {$e^+$};
                \vertex [below left=0.6cm and 1.5cm of v2] (i2) {$e^-$};
                \vertex [below right=0.6cm and 1.5cm of v2] (f2) {$e^-$};
            
                \diagram* {
                  (i1) -- [anti fermion] (v1) -- [anti fermion] (f1),
                  (i2) -- [fermion] (v2) -- [fermion] (f2),
                  (v1) -- [boson, edge label=$Z$] (vH),
                  (v2) -- [boson, edge label'=$Z$] (vH),
                  (vH) -- [scalar] (fH),
                };
              \end{feynman}
            \end{tikzpicture}
            \caption{VBF ($e^+ e^- \to e^+ e^- h$).}
            \label{fig:VBF_sub}
        \end{subfigure}
        
        \caption{Higgs production modes considered in this analysis.}
        \label{fig:ZH_VBF_combined}
    \end{figure}
}

\NewDocumentCommand{\leptonCombinationsTable}{}{
  \begin{table}[!ht]
      \centering
      \begin{ruledtabular}
      \begin{tabular}{cccc}
      $\ell' (1^{\text{st}})$ & $\ell (2^{\text{nd}})$ & $\ell (3^{\text{rd}})$ & $\ell (4^{\text{th}})$ \\
      \hline
      $\mu^+$ & $e^+$ & $e^-$ & $e^-$ \\
      $\mu^-$ & $e^-$ & $e^+$ & $e^+$ \\
      $e^+$ & $\mu^+$ & $\mu^-$ & $\mu^-$ \\
      $e^-$ & $\mu^-$ & $\mu^+$ & $\mu^+$ \\
      \end{tabular}
      \end{ruledtabular}
      \caption{All four-lepton combinations.}
      \label{tab:lepton_combinations}
  \end{table}
}

%% file: abstract.tex
We investigate the projected sensitivity of the FCC-ee in probing the LFV decay of the 125 GeV Higgs boson,
$h\to\mu\tau$, $h\to e\tau$, and $h \to e\mu$ at center-of-mass energy of 240 GeV with an integrated luminosity of 5 ab$^{-1}$. 
With the clean multi-lepton final states, we consider the Higgs production modes that contribute to $e^+e^- \to \ell\ell h$,
in particular, the Higgs-strahlung and the vector boson fusion production channels. The 95\% CL upper limits on branching ratio are found to be $5.94 \times 10^{-4}$, $6.28 \times 10^{-4}$, and $7.37 \times 10^{-5}$ for the channels $h\to\mu\tau$, $h\to e\tau$, and $h \to e\mu$, respectively. In addition, we also consider the LFV decay of a new CP-even Higgs boson with mass in the range of $110-200$ GeV. The projected FCC-ee sensitivities are then compared to those from low energy searches for $\ell\to\ell'\gamma$ decay. We find that the FCC-ee sensitivity for the LFV decays of the Higgs in the $e-\tau$ and $\mu-\tau$ channels is superior to the low-energy constraints. However, in the $e-\mu$ case, the low energy search offers stronger constraints.


%% file: introduction.tex
\section{Introduction}

The 2012 discovery of the 125 GeV Higgs boson ($h$) jointly by the ATLAS~\cite{ATLAS:2012yve} and CMS~\cite{CMS:2012qbp} collaborations was a breakthrough in particle physics. It completes the Standard Model (SM) of particle physics, which describes elementary particles and their interactions. Even though the $h$ properties agree very well with the SM predictions~\cite{Eur.Phys.J.C76.2016.6,Eur.Phys.J.C.75.2015.212, Phys.Rev.D111.2025.092014}, there are reasons to expect new physics beyond the SM (BSM). Chief indications among them are the massive nature of neutrinos, the existence of dark matter, and the baryon-antibaryon asymmetry of the Universe. 
These reasons motivate searches for BSM physics, with the Higgs sector offering the possibility to reveal new interactions. 

Lepton flavor violating (LFV) decays of the 125 GeV Higgs boson, $h\to e\mu$, $h\to e\tau$, $h\to \mu\tau$, are forbidden in the 
SM. These decays, if observed, would provide clear evidence for BSM physics. In addition, the LFV decays of $h$ are expected in several BSM frameworks, including the two-Higgs-doublet-model (2HDM)~\cite{Branco:2011iw,Arroyo-Urena:2013cyf,Crivellin:2013wna,Altmannshofer:2016zrn,Arroyo-Urena:2020mgg,Arroyo-Urena:2023vfh}, 
models with extra-dimension~\cite{Perez:2008ee,Casagrande:2008hr,Albrecht:2009xr}, models with supersymmetry~\cite{Diaz_Cruz_2000,Han:2000jz,Arhrib:2012ax,Arana-Catania:2013xma,Arganda:2015naa,Arganda:2015uca} 
, models for neutrino mass generation~\cite{PILAFTSIS199268,Arganda:2004bz,Arganda:2014dta,Arganda:2016zvc,Marcano:2019rmk,Primulando:2022lpj,Primulando:2022vip}, flavor models with Froggatt-Nielsen mechanism~\cite{Arroyo-Urena:2025gcs,Arroyo-Urena:2018mvl}, and model with leptoquarks~\cite{Urena:2021xtw}.
Moreover, additional Higgs bosons ($H$) in BSM models can generically accommodate LFV decays. The LFV decay modes of both the $h$ and $H$ offer a unique opportunity to probe BSM physics at particle colliders such as the Large Hadron Collider (LHC)~\cite{Primulando:2016eod,Primulando:2019ydt,Arganda:2019gnv,Barman:2022iwj,Primulando:2023ugc,Afik:2023vyl}. 

Existing LHC searches have already placed strong constraints on the LFV decays of the 125 GeV Higgs boson. The CMS collaboration has set 95\%~confidence level (CL) upper limits of $\mathcal{B}(h\to e\mu) < 4.4\times10^{-5}$~\cite{CMS:2023pte}, $\mathcal{B}(h\to\mu\tau) < 0.15\%$~\cite{PhysRevD.104.032013}, and $\mathcal{B}(h \to e\tau) < 0.22\%$~\cite{PhysRevD.104.032013}. The ATLAS collaboration has provided slightly stronger upper limits $\mathcal{B}(h\to e\mu) < 6.2\times10^{-5}$~\cite{ATLAS:2019old}, $\mathcal{B}(h\to\mu\tau) < 0.09\%$~\cite{JHEP.07.2023.166}, and $\mathcal{B}(h\to e\tau) < 0.12\%$~\cite{JHEP.07.2023.166}. In addition, both CMS and ATLAS have also searched for LFV decays of the additional Higgs boson $H$. The $H\to e\mu$ decay mode has been investigated by both CMS and ATLAS in the mass range 110 GeV $\le m_H\le 160$~\cite{ATLAS:2019old,CMS:2023pte}. In particular, CMS reports a potential excess at $m_H=146$ GeV with a local (global) significance of 3.8 (2.8) standard deviations ($\sigma$). The $H\to e\tau$ and $H\to\mu\tau$ decays have been searched for by CMS in the mass range of $200-900$ GeV with no significant excess reported~\cite{CMS:2019pex}. Additionally, the $H\to \mu\tau$ mode has also been investigated by LHCb in the mass range of $45-190$ GeV~\cite{LHCb:2018ukt}. 

\ZHVBFdiagram

The planned Future Circular Collider in $e^+e^-$ mode (FCC-ee) offers a clean environment for studying the Higgs LFV decays. At the operating center-of-mass energy of 240 GeV, both the 125 GeV Higgs boson and the additional Higgs boson can be produced via the Higgs-strahlung process, shown in Figure~\ref{fig:ZH_sub}. This makes it possible to probe LFV decays in either $q\bar q\ell\ell'$ or $\ell\bar\ell\ell\ell'$ (4-lepton) final states. Note here that $\ell=e,\mu$, and $\tau$ is decays leptonically. Ref.~\cite{Qin_2018}, taking advantage of the larger $Z\to q\bar q$ branching ratio, has investigated the FCC-ee reach in the $q\bar q\ell\ell'$ final states for a 125 GeV Higgs boson. Even though the 4-lepton final states have a lower cross-section for the channel with $Z\to e^+e^-$, they get an additional contribution from the $Z$-boson fusion process, as shown in Figure~\ref{fig:VBF_sub}. This extra contribution becomes important for $m_H\gtrsim150$ GeV, where the $Z$ boson in the Higgs-strahlung process is off-shell.
In this paper, we study the sensitivity of the FCC-ee to the LFV decays of the Higgs boson. We focus our analysis on the 4-lepton final states, which allow us to probe additional Higgs boson masses up to 200 GeV. While not advocating any particular BSM scenario, we interpret our results in the context of the type-III 2HDM. This allows us to compare our FCC-ee projections to low energy LFV constraints from $\mu\to e\gamma$ and $\tau\to \ell\gamma$ decays. 

This paper is organized as follows: Section~\ref{sec:mc_simulation} describes the simulation of signal and background processes. Section~\ref{sec:event_reconstruction} details the reconstruction of the LFV Higgs boson, including collinear mass reconstruction for LFV Higgs decays with $\tau$ leptons. The event selection of the three LFV channels, divided into two phase space regions based on the expected mass of the associated lepton pair, is described in Section~\ref{sec:event_selection}. In Section~\ref{sec:results}, we present the FCC-ee projected sensitivities to the LFV Higgs decays with an integrated luminosity of 5 ab$^{-1}$. The obtained sensitivities are then interpreted in the context of the type-III 2HDM and compared to current and future low energy LFV constraints in Section~\ref{sec:model}. Finally, all findings are summarized in Section~\ref{sec:summary}.

%% file: simulation.tex
\section{Signal and Background}
\label{sec:mc_simulation}

The hard scattering part of both the signal and background processes is generated using MadGraph5\_aMC@NLO~\cite{MadGraph} at $\sqrt{s}=240$ GeV. We then use Pythia8~\cite{Pythia} to handle the final state radiation, the decay of the $\tau$ leptons, and the LFV decays of the Higgs bosons. Finally, the detector response is simulated using Delphes~\cite{Delphes} with the IDEA detector card to evaluate the reach of FCC-ee following the design parameters described in Ref.~\cite{IDEA}. 
To reflect the realistic conditions of the collider, the effects of initial state radiation (ISR) and beamstrahlung are also included in the simulation to obtain accurate cross-sections, using ISR and beamstrahlung parameterizations for the FCC-ee scenario at $\sqrt{s}=240$ GeV~\cite{Frixione:2021zdp}.
The FCC-ee detector is expected to be capable of tracking very soft leptons, with 98\% efficiency down to 100 MeV in transverse momentum ($\pt$), and pseudo-rapidity ($\eta$) coverage up to 2.6~\cite{barchetta2021trackingvertexdetectorsfccee}. However, to conservatively estimate detector performance, we impose the following generator-level cuts in MadGraph: lepton $\pt > 1$ GeV, $|\eta| < 2.5$ for all leptons, and a separation requirement of $\Delta R > 0.4$ between the leptons. 

\subsection{Signal Processes}
In this paper, we focus on the final state with 4 light leptons, $\ell\ell\ell\ell'$, where the LFV pair $\ell\ell'$ originates either from the 125 GeV Higgs, $h$, or the hypothetical CP-even scalar, $H$. The additional lepton pair $\ell\ell$ either comes from the $Z$ boson decay in the Higgs-strahlung ($ZH$) production mode or from the scattered leptons in the VBF production mode. In our analysis, we consider the inclusive process $e^+ e^- \to \ell^+\ell^- H$, containing both $ZH$ and VBF contributions, as well as their interference effects. To simplify our analysis, we assume the coupling of $H$ to the $Z$-boson is SM-like. Treating the $H$ resonance mass, $m_H$, as a free parameter, we study 100 GeV $\le m_H\le200$ GeV. Note that, with our assumption, the LFV signal of the $h$ is identical to that of $H$ when $m_H=125$ GeV. 

In our simulation, the signal events $e^+ e^- \to \ell^+\ell^- H$ are generated by MadGraph5 with the $H$ width set to zero, assuming the total width of the $H$ is negligibly small compared to its mass ($\Gamma_H / m_H \ll 1$) and falls well within the detector resolution. The $H$ is then forced to decay via Pythia8 into LFV channels: $\htomutaue$, $\htoetaumu$, and $\htomue$, where $\tau_\ell$ stands for $\tau\to\ell\nu_\ell\nu_\tau$. Note that the LFV signal processes involving $H\to\ell\tau$ also contain missing transverse energy ($\met$). In our analysis, we have generated 1 million events for each LFV signature. 

\begin{figure}[ht!]
    \centering
    \includegraphics[width=0.45\textwidth]{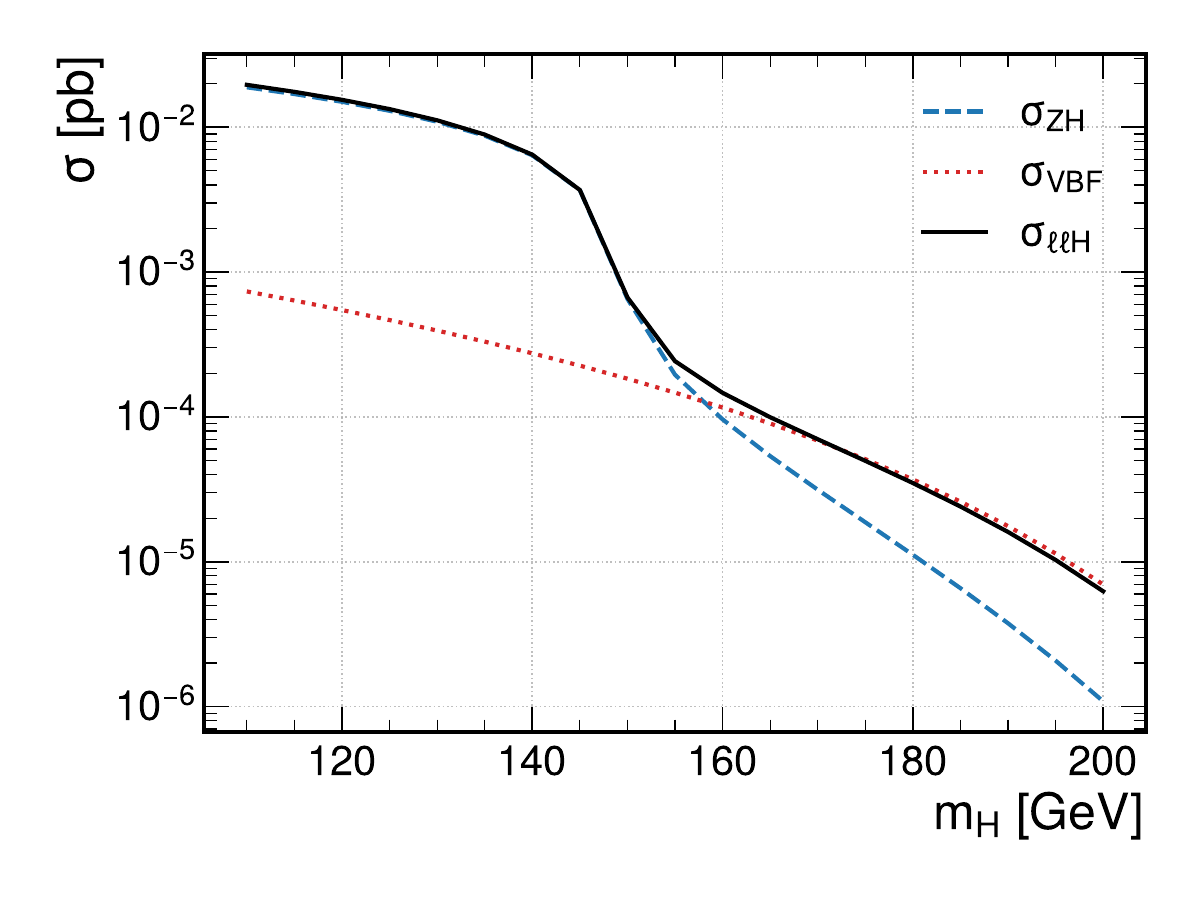}
    \caption{Cross sections of the inclusive $e^+ e^- \to \ell^+\ell^- H$ (black), exclusive $ZH$ with $Z \to \ell\ell$ (blue), and VBF (red) processes as a function of $m_H$. The cross sections are calculated at LO using MadGraph5.}
    \label{fig:cross_section_comparison}
\end{figure}

The cross-sections for signal processes at leading order (LO) across the mass range $110-200$  GeV are shown in Table~\ref{tab:signal_cross_sections}. Separate $ZH$ and VBF samples were generated to illustrate their relative contributions. As shown in Fig.~\ref{fig:cross_section_comparison}, the relative contribution from VBF becomes increasingly important above $m_H \approx 150$ GeV due to the suppression of on-shell $ZH$ near the kinematic limit.

\begin{table}[h!]
    \centering
    \begin{ruledtabular}
    \begin{tabular}{cccc}
        & \multicolumn{3}{c}{Cross section (fb)} \\
        $m_H$ (GeV) & $ZH, Z \to \ell\ell$ & VBF & $\ell\ell H$ \\ \midrule
        110 & 18.87 & 0.735 & 19.63 \\
        120 & 14.99 & 0.547 & 15.47 \\
        125 & 12.99 & 0.466 & 13.35 \\
        130 & 10.92 & 0.395 & 11.17 \\
        140 & 6.401 & 0.275 & 6.471 \\
        150 & 0.653 & 0.184 & 0.669 \\
        160 & 9.62$\times 10^{-02}$ & 0.116 & 0.147 \\
        170 & 3.14$\times 10^{-02}$ & 6.86$\times 10^{-02}$ & 7.01$\times 10^{-02}$ \\
        180 & 1.12$\times 10^{-02}$ & 3.70$\times 10^{-02}$ & 3.50$\times 10^{-02}$ \\
        190 & 3.76$\times 10^{-03}$ & 1.76$\times 10^{-02}$ & 1.61$\times 10^{-02}$ \\
        200 & 1.09$\times 10^{-03}$ & 6.96$\times 10^{-03}$ & 6.25$\times 10^{-03}$ \\
    \end{tabular}
    \end{ruledtabular}
    \caption{Cross-sections of the signal processes. The inclusive $\ell\ell H$ process includes both $ZH$ and VBF contributions, accounting for interference effects, while the exclusive $ZH$ and VBF processes are generated separately. The cross sections are calculated at LO using MadGraph5. 
    }
    \label{tab:signal_cross_sections}
\end{table}

\subsection{SM Backgrounds}
This study focuses on irreducible SM backgrounds yielding four-lepton final states. Reducible backgrounds involving jet or photon misidentification are expected to be strongly suppressed and are therefore not considered in this analysis.
{
The SM background originates from $ZZ$ production, which contributes to the signal region via the $ZZ \to \ell\ell\tau\tau$ process, where the $\tau$ leptons decay leptonically to produce the four-lepton signatures.  
Moreover, there are contributions from the Higgs-strahlung ($Zh$) production mode where $Z \to \ell\ell$ and $h \to \tau\tau$. Additional background arises from electroweak processes (EW) $\ell^+\ell^-W^{+(*)}W^{-(*)}$ where both $W^{(*)}$ decay to light lepton.
}

The SM background processes considered in this analysis are summarized as follows:
\begin{itemize}
    \item {$Zh$ with $Z \to \ell\ell$ and $h \to \tau\tau$}
    \item {$ZZ \to \ell\ell\tau\tau$}
    \item {EW ($\ell^+\ell^-W^{+(*)}W^{-(*)}$) with $WW^* \to \ell \ell \nu\bar{\nu}$}
\end{itemize}
We have generated 10 million events for each background category. The cross-sections for these background processes calculated by MadGraph5 at LO are listed in Table~\ref{tab:background_cross_sections}.

\begin{table}[th!]
    \centering
    \begin{ruledtabular}
    \begin{tabular}{lc}
        Process & Cross section (fb) \\ \midrule
        $ZZ \to \ell\ell\tau\tau$ & 5.02 \\
        $Zh$ ($Z\to\ell\ell$, $h \to \tau\tau$) & 0.53 \\
        {$\ell^+\ell^-W^{+(*)}W^{-(*)}$, $WW^* \to \ell \ell \nu\bar{\nu}$} & {0.23}
    \end{tabular}
    \end{ruledtabular}
    \caption{Cross sections of the background processes.}
    \label{tab:background_cross_sections}
\end{table}

%% file: event_reconstruction.tex
\section{Event Reconstruction}
\label{sec:event_reconstruction}
\subsection{Lepton Reconstruction and Object Definitions}

Following the IDEA detector setup, reconstruction for both $e$ and $\mu$ is assumed to be perfect at $p_T > 100$ MeV and $|\eta| \leq 2.5$. The $\tau$ lepton reconstruction is based on the visible decay products. Note that the lepton transverse momentum is still bounded by $p_T > 1$ GeV from the generator cut.

The missing transverse energy 
is reconstructed as the negative vector sum of the transverse momenta of all reconstructed particles in the event. Any particles outside the detector acceptance or below the reconstruction thresholds are not included in this sum, leading to an additional contribution to the $\met$ from those undetected particles. 

\subsection{Higgs Mass Reconstruction}
\label{subsec:col_mass}

To reconstruct the Higgs mass from its decay products, one needs to know the energy and momentum of all the daughters. However, in the case of $\htoetaumu$ and $\htomutaue$, this is impossible since the leptonically decayed $\tau$ produces two neutrinos. In this case, we instead reconstruct the collinear mass. 
%
Under the collinear approximation, the neutrinos from the $\tau$ decay are
assumed to be emitted in the same direction as the visible $\tau$ decay
products. The visible momentum fraction is defined as

\begin{equation}
x_{\tau}^{\text{vis}} =
\frac{p_{T}^{\tau,\text{vis}}}
     {p_{T}^{\tau,\text{vis}} + p_{T}^{\nu,\text{est}} },
\end{equation}
where $p_{T}^{\nu,\text{est}}$ is the component of $\vec{p}_T^{ \text{miss}}$
projected onto the direction of the visible $\tau$ candidate. This can be written as
\begin{equation}
p_{T}^{\nu,\text{est}} = \vec{p}_T^{ \text{miss}} \cdot \hat{p}_T^{\tau,\text{vis}},
\end{equation}
where $\hat{p}_T^{\tau,\text{vis}}$ is the unit vector in the direction of the visible $\tau$ candidate's transverse momentum.

The collinear mass is then reconstructed from the visible invariant mass
$M_{\text{vis}}$ of the $\tau\text{--}\ell$ system (where $\ell = e$ or $\mu$) as
\begin{equation}
M_{\text{col}} = \frac{M_{\text{vis}}}{\sqrt{x_{\tau}^{\text{vis}}}}.
\label{eq:collinear_mass}
\end{equation}
This definition of collinear mass was first proposed in Ref.~\cite{Elagin_2011} and was also used in the previous CMS analysis for the same Higgs LFV decay final states in $pp$ collisions at 13 TeV~\cite{PhysRevD.104.032013}.

For the $\htomue$ channel, the Higgs candidate mass can be reconstructed directly from the invariant mass of the two leptons as
\begin{equation}
M_{\ell\ell} = \sqrt{(E_{\ell_1} + E_{\ell_2})^2 - |\vec{p}_{\ell_1} + \vec{p}_{\ell_2}|^2}.
\label{eq:invariant_mass}
\end{equation}

%% file: event_selections.tex
\section{Event Selection}
\label{sec:event_selection}

In this analysis, we focus on three LFV channels: $\htomutaue$, $\htoetaumu$, and $\htomue$. In both the $ZH$ and VBF production modes, the final state is characterized by a pair of opposite-sign and same flavor (OSSF) light leptons ($\ell\bar\ell$) produced in association with the $H$. In the $ZH$ production mode, these associated leptons originate from the $Z$ boson decay, whereas in the VBF, they represent the scattered leptons. An additional opposite-sign and different flavor (OSDF) lepton pair ($\ell\ell’$) in the final state arises from the LFV $H$ decay. This results in a four-lepton final state in an asymmetric flavor combination, either $\mu eee$ or $e\mu\mu\mu$ together with $\met$ from $\tau$ decay exclusively in channels $\htomutaue$ and $\htoetaumu$. To optimize sensitivity across all mass points for both $ZH$- and VBF-dominant regions, multiple selection categories are considered, where each category is optimized for an individual production mode. These are discussed in Subsection~\ref{subsec:z_boson_reconstruction}. As three LFV channels share a similar final state topology except for $\met$ in channel $\htomutaue$ and $\htoetaumu$, a common preselection and candidate identification are applied to all three channels, followed by specific selections optimized for each, as detailed in Subsections~\ref{subsec:specific_selection_met} and~\ref{subsec:specific_selection_no_met}.

\subsection{Common Preselection}
\label{subsec:preselection}

The initial selection requires exactly four leptons with $\pt > 1$ GeV, either in the combinations of $\mu eee$ or $e\mu\mu\mu$ with charge conservation. Events with additional lepton candidates are discarded. Following the combinatorial logic, all possible four-lepton combinations are listed in Table~\ref{tab:lepton_combinations}.

\leptonCombinationsTable

From the combinations $3\ell + 1\ell'$, the lone lepton ($\ell'$) is identified as the $H$ candidate denoted as $\ell_H$. Then, the same-sign and different flavor (SSDF) lepton to $\ell'$ is identified as the primary associated lepton ($\ell_A$). This assignment is based on the fact that this lepton cannot originate from the same vertex as $\ell_H$ in an LFV decay without violating charge conservation. The two remaining leptons of the OSSF are treated as ambiguous candidates for the associated pair.
\subsection{Candidate Identification}
\label{subsec:z_boson_reconstruction}

To reconstruct the associated leptons originating either from the $Z$ boson decay in $ZH$ or from the scattering process in VBF, the identified $\ell_A$ in the preselection step is paired with one of the two ambiguous leptons. This results in two possible pairs.

For the $ZH$ production mode, the invariant mass of the lepton pair from $Z$ is expected to peak at the $Z$ mass of 91 GeV if the $Z$ boson is produced on-shell. For the off-shell $Z$-boson, the distribution extends toward lower masses, with the upper bound determined by the phase space limit at $\sqrt{s}-m_H$. Conversely, the associated leptons in the VBF production mode originate from the non-resonant scattering electrons pair. This results in a broad continuum in the invariant mass of the associated lepton pair ($M(\ell\ell_A)$) distribution that also extends toward the lower mass region, while the upper bound is subject to the same phase space constraints as in the $ZH$ case.

Events are classified into two orthogonal categories based on $M(\ell\ell_A)$:
\begin{itemize}
    \item \textbf{$Z$-mass category}: Requires $M(\ell\ell_A) \in [81, 101]$ GeV. If multiple pairs satisfy this, the pair with the mass closest to the $Z$-boson mass is selected.
    \item \textbf{Low-mass category}: Requires $M(\ell\ell_A) \in [21, 81]$ GeV to include contributions from off-shell $ZH$ and soft associated leptons in VBF. In the case of multiple pairs, the one with the highest $M(\ell\ell_A)$ is chosen. This is motivated by the fact that invariant mass distributions for off-shell $ZH$ and VBF processes typically increase toward the phase space limit at $\sqrt{s}-m_H$. A lower bound of 21 GeV is applied to suppress the contribution from backgrounds with a photon as a propagator and low-mass resonance.
\end{itemize}

To preserve orthogonality between the two categories and avoid double-counting, any events containing a lepton pair that satisfies the $Z$-mass criteria is automatically categorized as $Z$-mass category, and only the remaining events are considered for the low-mass selection.
Once the associated lepton pair is identified, the remaining lepton in the event is automatically assigned to the $H$ candidate.

\subsection{$\htomutaue$ and $\htoetaumu$}
\label{subsec:specific_selection_met}

The final state from $H$ LFV decay consists of one prompt lepton $\ell_\text{prompt}$, one visible decay product of $\tau$, denoted as $\tau_\textrm{vis}$, and $\met$ from the neutrinos produced in the $\tau$ decay. The prompt lepton from $H$ decay is expected to be energetic, whereas $\tau_\textrm{vis}$ is expected to be softer. Therefore, the prompt lepton is required to satisfy $\pt(\ell_\text{prompt}) > 10$ GeV.

\begin{figure*}[tbp]
    \centering
    \begin{subfigure}{0.99\textwidth}
        \includegraphics[width=\textwidth]{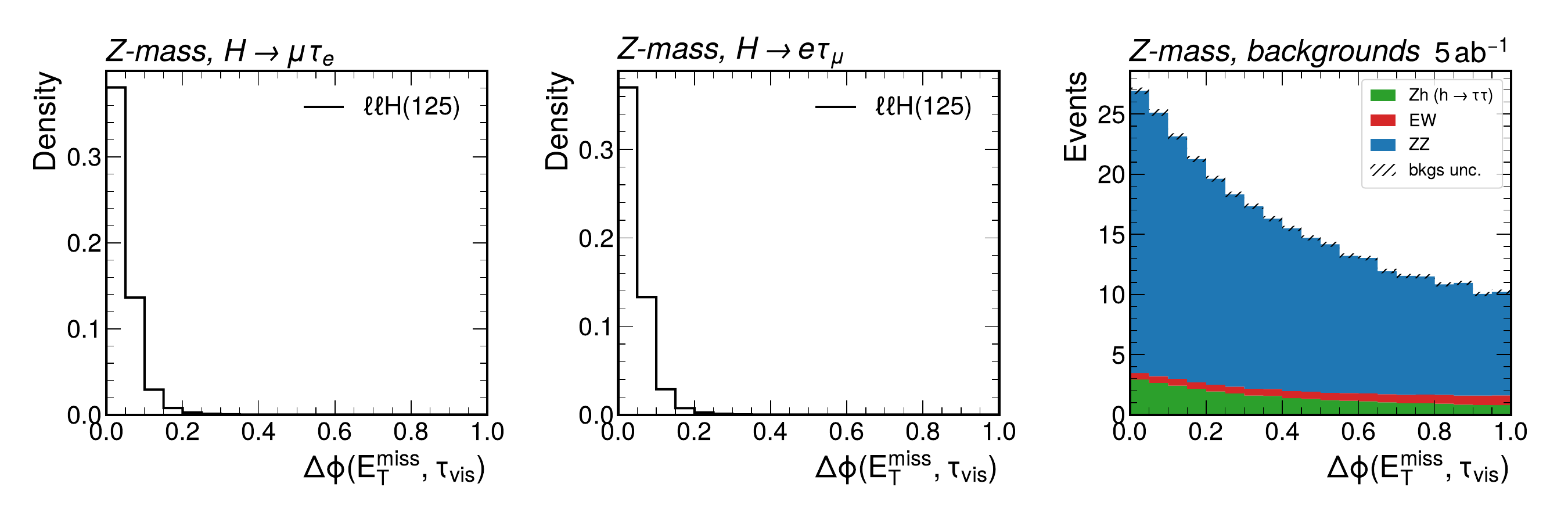}
    \end{subfigure}
    \hfill
    \caption{Distributions of $\Delta \phi(\met, \tau_\textrm{vis})$ after the candidate identification for signal for the $\htomutaue$ and $\htoetaumu$ channels at $m_H = 125$ GeV in $Z$-mass region. The right figure shows the background normalized to 5 ab$^{-1}$.}
    \label{fig:dphi_met_visTau}
\end{figure*}

The missing energy from the leptonic decay of the $\tau$ is required to be $\met>10$ GeV. Additionally, neutrinos from $\tau$ decay products are expected to be highly boosted in the same direction as $\tau_\textrm{vis}$. To exploit this signal characteristic, an additional requirement is imposed on the azimuthal angle difference between $\met$ and $\tau_\textrm{vis}$, requiring that $\Delta \phi(\met, \tau_\textrm{vis}) < 0.1$. This threshold was optimized by minimizing the expected upper limit, as it suppresses the majority of backgrounds while retaining $>90\%$ signal. The distributions of $\Delta \phi(\met, \tau_\textrm{vis})$ are illustrated in Figure~\ref{fig:dphi_met_visTau}. 


\subsection{$\htomue$}
\label{subsec:specific_selection_no_met}

Unlike the $\htomutaue$ and $\htoetaumu$ cases, the $\htomue$ channel consists of two prompt leptons without neutrinos from the Higgs decay. Both leptons are then expected to be energetic, leading to the requirement that $\pt(e) > 10$ GeV and $\pt(\mu) > 10$ GeV.

\begin{figure*}[tbp]
    \centering
    \begin{subfigure}{0.99\textwidth}
        \includegraphics[width=\textwidth]{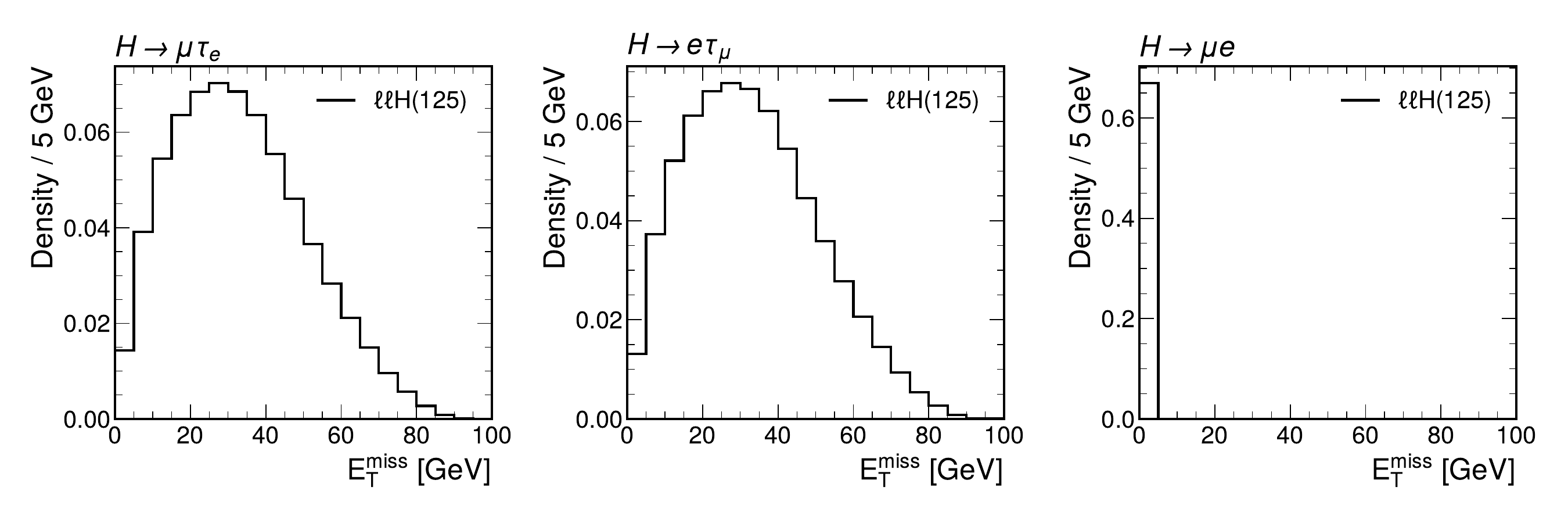}
        \caption{signals.}
    \end{subfigure}
    \hfill
    \begin{subfigure}{0.33\textwidth}
        \includegraphics[width=\textwidth]{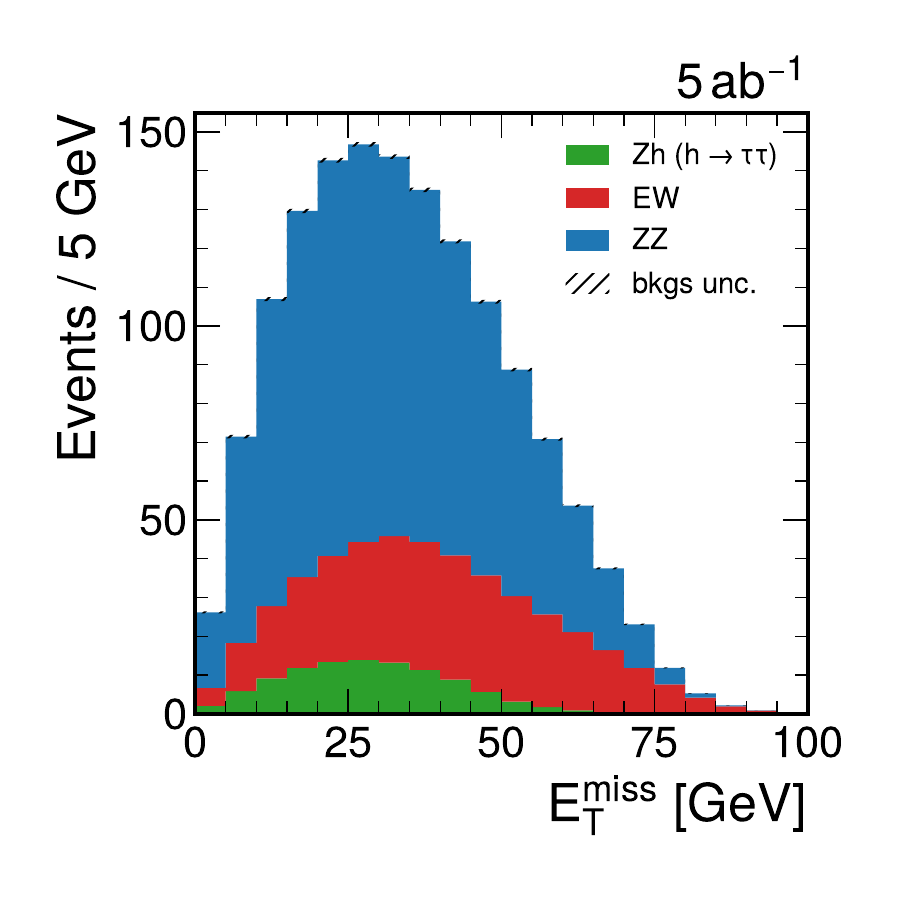}
        \caption{backgrounds.}
    \end{subfigure}
    \caption{Preselection distributions of $\met$ for a) signal at $m_H = 125$ GeV and b) background normalized to 5 ab$^{-1}$. Notice the absence of $\met$ in the $e\mu$ channel due to the final state not providing any invisible neutrinos.}
    \label{fig:met_distribution}
\end{figure*}

Due to the absence of neutrinos in the signal, 
we require $\met < 10$ GeV. Distributions of $\met$ at the pre-selection level for three channels and backgrounds are presented in Figure~\ref{fig:met_distribution}. Additionally, Table~\ref{tab:cutsummary} summarizes the cuts on $\pt(\ell)$, $\met$, and $\Delta \phi(\met, \tau_\textrm{vis})$ for all the 3 LFV channels. 

\begin{table}[hp]
    \centering
    \begin{ruledtabular}
    \begin{tabular}{lccc}
        Selection & $H \to \mu \tau_e$ & $H \to e \tau_\mu$ & $H \to \mu e$ \\
        \midrule
        $p_T(\mu)$ & $> 10$ GeV & - & $> 10$ GeV \\
        $p_T(e)$ & - & $> 10$ GeV & $> 10$ GeV \\
        $\met$ & $> 10$ GeV & $> 10$ GeV & $< 10$ GeV \\
        $\Delta \phi(\met, \tau_\textrm{vis})$ & $< 0.1$ & $< 0.1$ & - \\
    \end{tabular}%
    \end{ruledtabular}
    \caption{Selection criteria for three channels.
    }
    \label{tab:cutsummary}
\end{table}


%% file: results.tex
\section{Results}
\label{sec:results}

\begin{figure*}[tpb]
    \centering
    \includegraphics[width=0.99\textwidth]{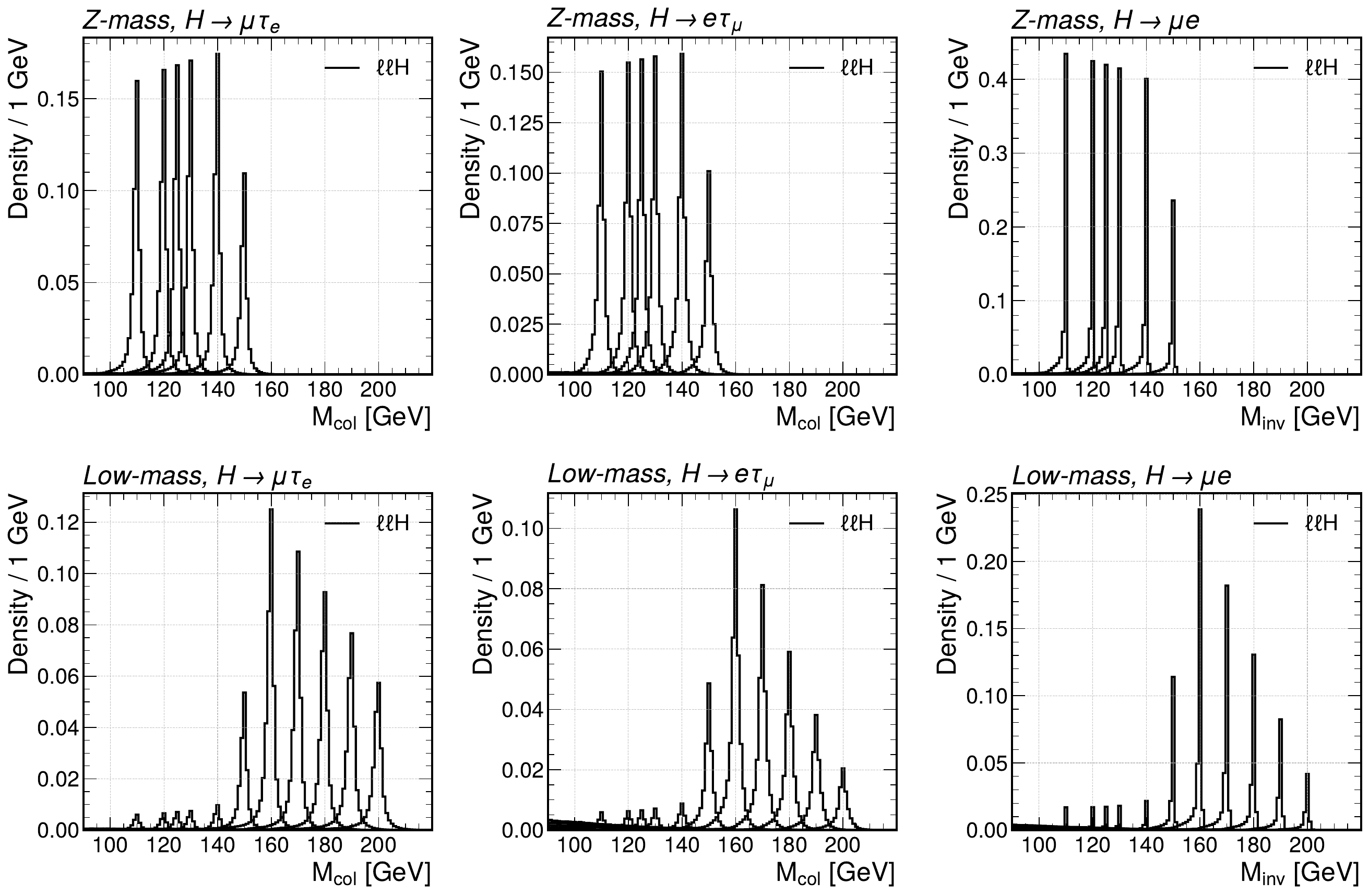}
    \caption{Reconstructed $H$ mass distributions for three channels after all selections. Signals shown include $m_H =110-200$  GeV with 10 GeV increments, in addition to 125 GeV. Notice the absence of signals at $m_H > 150$ GeV in $Z$-mass categories and small contributions from the VBF production mode in the low-mass category at $m_H < 150$ GeV.}
    \label{fig:reconstruction_performance}
\end{figure*}

After all selections are applied, $H$ is then reconstructed using the collinear mass method for channels involving $\tau$ leptons and the traditional invariant mass method for the $\htomue$ channel. The prefit distributions of the reconstructed mass are displayed in Figure~\ref{fig:reconstruction_performance}. In the $Z$-mass region, the analysis targets signals with on-shell $Z$ boson decays. However, reconstructing $H$ in this category is precluded for the signal with $m_H > 150$ GeV, as the allowed mass for the associated leptons is constrained by the phase space limit $M(\ell\ell) \leq \sqrt{s} - m_H$. Consequently, the low-mass category is designed specifically to target heavy $H$ in the $ZH$ and the VBF production modes. Note that in the transition mass range of $150 \leq m_H < 160$ GeV, the signal yields for both categories are significant.

\begin{figure*}[tpb]
    \centering
    \includegraphics[width=0.99\textwidth]{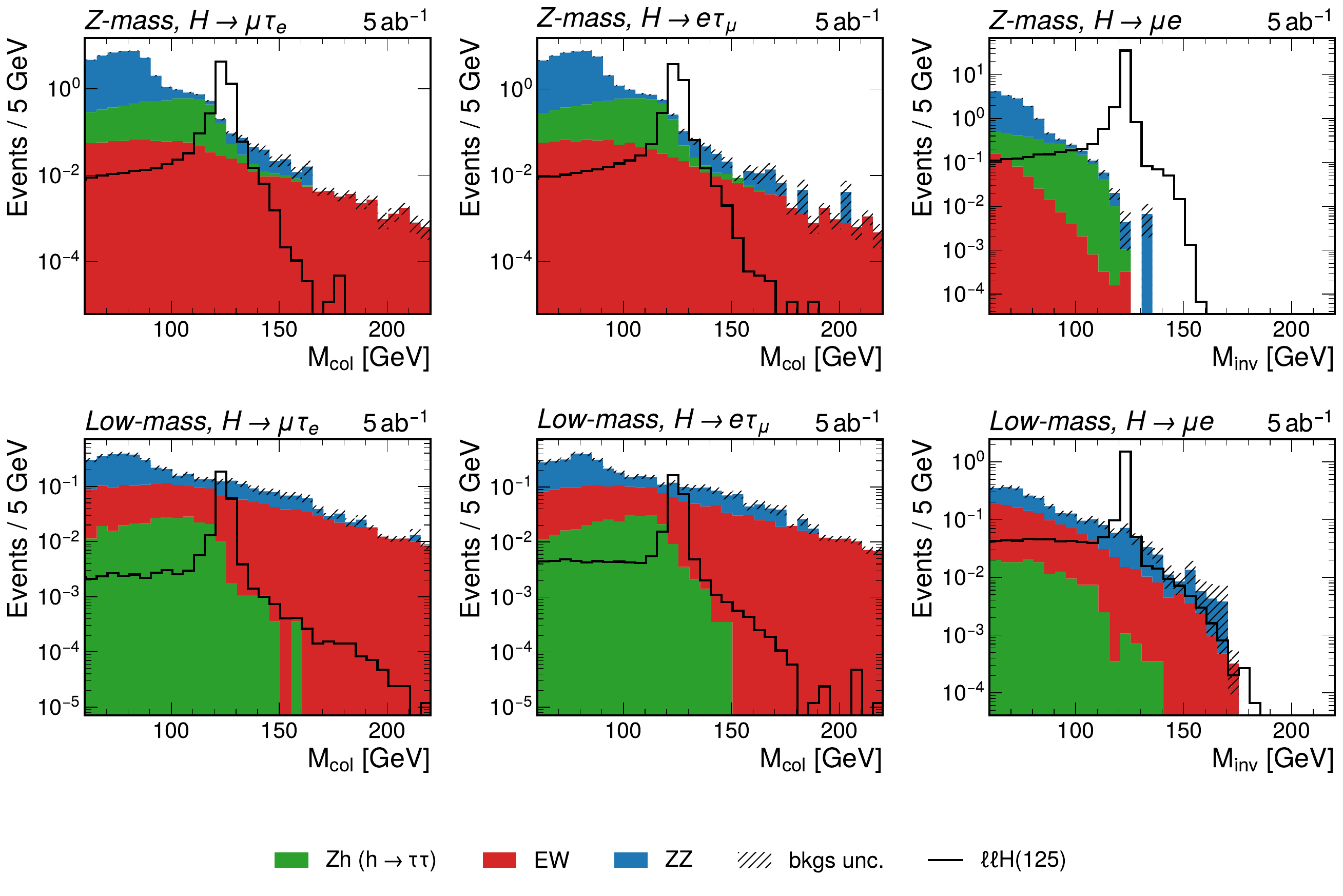}
    \caption{Distributions of reconstructed $H$ mass for both signal and background processes after all selections. The signal shown is for $m_H = 125$ GeV only.}
    \label{fig:reconstruction_all}
\end{figure*}

Figure~\ref{fig:reconstruction_all} presents the distributions of the reconstructed $H$ for both the expected backgrounds and the signals after all selections. The signal yields are normalized to $\sigma_{\ell\ell H} \times \mathcal{L} \times \mathcal{B}(H \to \ell \ell')$, assuming an integrated luminosity of $\mathcal{L} = 5$ ab$^{-1}$ and a branching ratio of $\mathcal{B}(H \to \ell \ell') = 10^{-3}$, with inclusive production cross-section $\sigma_{\ell \ell H}$ listed in Table~\ref{tab:signal_cross_sections}. For the $\htomutaue$ and $\htoetaumu$ channels, an additional factor of $\mathcal{B}(\tau \to \ell \nu \bar{\nu}) \approx 0.17$ is applied to account for the leptonic tau decays. Similarly, all SM backgrounds are normalized to 5 ab$^{-1}$ using the cross-section provided in Table~\ref{tab:background_cross_sections}. Note that both the integrated luminosity and the collision energy are chosen to align with the $Zh$ production program at FCC-ee, planned for a three-year operation~\cite{FCCeePlan}.

To mitigate the impact of backgrounds, only the signal and backgrounds around the mass point of interest are considered for the upper limit calculation. The mass window is defined as $|M_\text{reco} - m_H| \leq 10$ GeV, where $M_\text{reco}$ is the collinear mass ($M_\text{col}$) for $\htomutaue$ and $\htoetaumu$ channels, and the invariant mass ($M_\text{inv}$) for $\htomue$ channel. The signal efficiencies and the number of events after all selections and the requirement $|M_\text{reco} - m_H| \leq 10$ GeV are summarized in Tables~\ref{tab:efficiency_yield_mutau}, \ref{tab:efficiency_yield_etau}, and~\ref{tab:efficiency_yield_mue} for $\htomutaue$, $\htoetaumu$, and $\htomue$, respectively.

For the $\htomutaue$ and $\htoetaumu$ channels, a dominant background arises from $Zh(h \to \tau\tau$) production due to similarities in their production modes. However, the collinear mass method effectively suppresses this background, as the presence of multiple sources of $\met$ makes it highly unlikely for the $\tau_{\text{vis}}$ to align with the $\met$ in the same way as the signal. As shown in Tables~\ref{tab:efficiency_yield_mutau} and \ref{tab:efficiency_yield_etau}, the $Zh(h \to \tau\tau)$ background in the $Z$-mass region yields approximately two events at $m_H =$ 110 GeV, but falls off dramatically to less than 0.1 events for $m_H \geq$ 140 GeV. The second largest contribution comes from $ZZ$ production. Although this process has the highest cross section among the SM backgrounds, its reconstructed mass tends to peak around the nominal $Z$ boson mass. This highly suppresses its contribution to the signal region. In the low-mass region, the prominent backgrounds are $ZZ$ (resulting from lepton mis-pairing) and EW processes. Despite this, the combined background remains exceptionally low, resulting in fewer than one expected observed event (peaking at roughly 0.6 events at 110 GeV) across all $m_H$.

The $\htomue$ channel provides the cleanest signal environment, as detailed in Table~\ref{tab:efficiency_yield_mue}. To mimic the fully visible $\ell\ell\ell\ell'$ final state, all relevant SM backgrounds must inherently contain intrinsic $\met$. Consequently, applying the $\met <$ 10 GeV condition highly suppresses these backgrounds. This reduces the expected number of background events to nearly zero across all mass points (with the highest 0.4 events at $m_H = 110$ GeV), while retaining $ZH$ signal efficiencies of up to 58\% in the $Z$-mass region.

The expected upper limits for LFV Higgs decays are calculated using the CMS Combine framework~\cite{CMS-Higgs-Combine}, employing the HybridNew method with the $\textrm{CL}_\textrm{s}$ criterion and Monte Carlo pseudo-experiments~\cite{CMS-NOTE-2011-005}. A joint likelihood is constructed from two counting experiments corresponding to the two event categories. The test statistic $\tilde{q}_\mu$ is defined using the profile likelihood ratio, with nuisance parameters profiled for each tested value of the signal strength $\mu$. For each value of $\mu$, distributions of $\tilde{q}_\mu$ are obtained from pseudo-data generated under the signal-plus-background and background-only hypotheses. The expected upper limit at 95\% confidence level is determined as the value of $\mu$ satisfying the $\textrm{CL}_\textrm{s}$ criterion of 0.05, based on the median of the test statistic $\tilde{q}_\mu$ distribution under the background-only hypothesis. The systematic uncertainties for all processes include the luminosity uncertainty of 2\%. Additionally, a conservative 10\% uncertainty is assigned to all backgrounds to account for all sources of systematic uncertainties (identification, reconstruction, and theoretical uncertainties). Note that the impact of these systematic uncertainties is small, as the search sensitivity is dominated by statistical uncertainties due to nearly zero expected observed events, especially for higher Higgs masses. For each LFV channel, the $Z$-mass and low-mass regions are combined to compute the expected upper limits on the LFV branching ratios $\mathcal{B}(\htomutau)$, $\mathcal{B}(\htoetau)$, and $\mathcal{B}(\htomue)$. The final limits are listed in Table~\ref{tab:limits_summary} together with the upper limits on the product $\sigma(e^+e^- \to \ell\ell H) \times \mathcal{B}(H \to \ell\ell')$ for each LFV channel as a function of $m_H$.

At $m_H = 125$ GeV, the expected upper limits on $\mathcal{B}(H \to \ell\ell')$ are found to be $5.94 \times 10^{-4}$, $6.28 \times 10^{-4}$, and $7.37 \times 10^{-5}$ for the $\htomutaue$, $\htoetaumu$, and $\htomue$ channels, respectively.  These results are complementary to the results obtained from the $ZH$ production mode with $Z \to q\bar{q}$ decay (equivalent to $e^+e^- \to q \Bar{q}H$) in Ref.~\cite{Qin_2018}, which can be combined with our results at $m_H = 125$ GeV to provide stronger constraints. Beyond this 125 GeV limit, the expected upper limits exhibit a mass-dependent trend across the probed $110-200$ GeV mass range, as summarized in Table~\ref{tab:limits_summary}. Generally, as the assumed $m_H$ increases, the upper limits on the branching ratios become less stringent. This behavior is driven by the fall in the total production cross-section, which drops from 19 fb at 110 GeV to merely 0.06 fb at 200 GeV. Consequently, the most stringent constraints across all three channels are achieved in the $110-130$  GeV $m_H$ region. As evident from the reconstructed mass in Figure~\ref{fig:reconstruction_performance} and the cross-section in Table~\ref{tab:signal_cross_sections}, the highest contribution at $m_H < 150$ GeV comes from the $ZH$ production mode, while the VBF production mode dominates at $m_H > 150$ GeV. As the assumed $H$ mass continues to increase, actual yields from the signal processes become lower than the expected upper limits of the signal event yields, given no observed events due to the vanishingly small production cross-section. This leads to the unphysical upper limits on the branching ratios (denoted by asterisks in Table~\ref{tab:limits_summary}), implying that the assumed integrated luminosity is simply insufficient to constrain the branching ratios in this regime. Notably, across the entire mass range, the $\htomue$ channel outperforms the $\htomutaue$ and $\htoetaumu$ channels by an order of magnitude due to its cleaner final state and the absence of $\met$ from tau decays, allowing for higher background suppression.

\begin{table*}[tpb]
    \centering
    \begin{ruledtabular}
    \begin{tabular}{ccccccccc }
        $m_H$ & \multicolumn{4}{c}{$Z$-mass Region} & \multicolumn{4}{c}{Low-mass Region} \\
        (GeV) & \multicolumn{1}{c}{Signal (Eff.)} & \multicolumn{3}{c}{Background (Evt.)} & \multicolumn{1}{c}{Signal (Eff.)} & \multicolumn{3}{c}{Background (Evt.)} \\
        & $\ell\ell H$ & $ZZ$ & $Zh(h\to\tau\tau)$ & EW & $\ell \ell H$ & $ZZ$ & $Zh(h\to\tau\tau)$ & EW \\
        \midrule
        110 & 0.47 & 0.85 & 2.04 & 0.20 & 0.02 & 0.21 & 0.10 & 0.30 \\
        120 & 0.48 & 0.35 & 1.15 & 0.14 & 0.02 & 0.21 & 0.06 & 0.27 \\
        125 & 0.49 & 0.21 & 0.62 & 0.11 & 0.02 & 0.21 & 0.04 & 0.25 \\
        130 & 0.50 & 0.15 & 0.21 & 0.09 & 0.02 & 0.21 & 0.02 & 0.23 \\
        140 & 0.50 & 0.11 & 0.02 & 0.05 & 0.03 & 0.16 & 0 & 0.20 \\
        150 & 0.32 & 0.06 & 0.01 & 0.04 & 0.17 & 0.13 & 0 & 0.17 \\
        160 & 0 & 0.03 & 0 & 0.03 & 0.40 & 0.10 & 0 & 0.14 \\
        170 & 0 & 0.01 & 0 & 0.02 & 0.37 & 0.06 & 0 & 0.11 \\
        180 & 0 & 0 & 0 & 0.01 & 0.33 & 0.02 & 0 & 0.09 \\
        190 & 0 & 0 & 0 & 0.01 & 0.29 & 0.01 & 0 & 0.07 \\
        200 & 0 & 0 & 0 & 0.01 & 0.23 & 0 & 0 & 0.05 \\
    \end{tabular}
    \end{ruledtabular}
    \caption{$\htomutaue$ selection efficiencies and background yields after all selections and $|M_\text{reco} - m_H| \leq 10$ GeV are applied. The background yields are normalized to $\mathcal{L} = 5$ ab$^{-1}$.}
    \label{tab:efficiency_yield_mutau}
    \end{table*}

\begin{table*}[tpb]
    \centering
    \begin{ruledtabular}
    \begin{tabular}{ccccccccc }
        $m_H$ & \multicolumn{4}{c}{$Z$-mass Region} & \multicolumn{4}{c}{Low-mass Region} \\
        (GeV) & \multicolumn{1}{c}{Signal (Eff.)} & \multicolumn{3}{c}{Background (Evt.)} & \multicolumn{1}{c}{Signal (Eff.)} & \multicolumn{3}{c}{Background (Evt.)} \\
        & $\ell\ell H$ & $ZZ$ & $Zh(h\to\tau\tau)$ & EW & $\ell \ell H$ & $ZZ$ & $Zh(h\to\tau\tau)$ & EW \\
        \midrule
        110 & 0.46 & 0.82 & 2.12 & 0.18 & 0.02 & 0.20 & 0.11 & 0.26 \\
        120 & 0.47 & 0.36 & 1.23 & 0.12 & 0.02 & 0.18 & 0.07 & 0.23 \\
        125 & 0.48 & 0.26 & 0.70 & 0.10 & 0.02 & 0.18 & 0.04 & 0.21 \\
        130 & 0.49 & 0.19 & 0.25 & 0.08 & 0.02 & 0.20 & 0.02 & 0.20 \\
        140 & 0.49 & 0.10 & 0.02 & 0.05 & 0.03 & 0.18 & 0 & 0.17 \\
        150 & 0.32 & 0.04 & 0.01 & 0.03 & 0.16 & 0.14 & 0 & 0.14 \\
        160 & 0 & 0.02 & 0 & 0.02 & 0.37 & 0.10 & 0 & 0.11 \\
        170 & 0 & 0.02 & 0 & 0.01 & 0.29 & 0.06 & 0 & 0.09 \\
        180 & 0 & 0.01 & 0 & 0.01 & 0.22 & 0.04 & 0 & 0.07 \\
        190 & 0 & 0 & 0 & 0.01 & 0.15 & 0.01 & 0 & 0.05 \\
        200 & 0 & 0 & 0 & 0 & 0.08 & 0 & 0 & 0.04 \\
    \end{tabular}
    \end{ruledtabular}
    \caption{$\htoetaumu$ selection efficiencies and background yields after all selections and $|M_\text{reco} - m_H| \leq 10$ GeV are applied. The background yields are normalized to $\mathcal{L} = 5$ ab$^{-1}$.}
    \label{tab:efficiency_yield_etau}

    \end{table*}
    
\begin{table*}[tpb]
    \centering
    \begin{ruledtabular}
    \begin{tabular}{ccccccccc }
        $m_H$ & \multicolumn{4}{c}{$Z$-mass Region} & \multicolumn{4}{c}{Low-mass Region} \\
        (GeV) & \multicolumn{1}{c}{Signal (Eff.)} & \multicolumn{3}{c}{Background (Evt.)} & \multicolumn{1}{c}{Signal (Eff.)} & \multicolumn{3}{c}{Background (Evt.)} \\
        & $\ell\ell H$ & $ZZ$ & $Zh(h\to\tau\tau)$ & EW & $\ell \ell H$ & $ZZ$ & $Zh(h\to\tau\tau)$ & EW \\
        \midrule
        110 & 0.58 & 0.09 & 0.30 & 0 & 0.03 & 0.18 & 0.02 & 0.14 \\
        120 & 0.58 & 0.03 & 0.06 & 0 & 0.02 & 0.18 & 0 & 0.08 \\
        125 & 0.58 & 0.02 & 0.01 & 0 & 0.02 & 0.15 & 0 & 0.06 \\
        130 & 0.58 & 0.01 & 0 & 0 & 0.03 & 0.13 & 0 & 0.05 \\
        140 & 0.56 & 0.01 & 0 & 0 & 0.03 & 0.05 & 0 & 0.03 \\
        150 & 0.34 & 0 & 0 & 0 & 0.17 & 0.02 & 0 & 0.02 \\
        160 & 0 & 0 & 0 & 0 & 0.36 & 0.02 & 0 & 0.01 \\
        170 & 0 & 0 & 0 & 0 & 0.28 & 0.01 & 0 & 0 \\
        180 & 0 & 0 & 0 & 0 & 0.21 & 0 & 0 & 0 \\
        190 & 0 & 0 & 0 & 0 & 0.14 & 0 & 0 & 0 \\
        200 & 0 & 0 & 0 & 0 & 0.07 & 0 & 0 & 0 \\
    \end{tabular}
    \end{ruledtabular}
    \caption{$\htomue$ selection efficiencies and background yields after all selections and $|M_\text{reco} - m_H| \leq 10$ GeV are applied. The background yields are normalized to $\mathcal{L} = 5$ ab$^{-1}$.}
    \label{tab:efficiency_yield_mue}
    \end{table*}

\begin{table*}[tph]
    \centering
    \begin{ruledtabular}
    \begin{tabular}{ ccccccc } 
        $m_H$ & \multicolumn{3}{c}{BR($H \to \ell\ell'$)} & \multicolumn{3}{c}{$\sigma(ee \to \ell\ell H) \times$ BR($H \to \ell\ell'$) (ab)} \\
        (GeV) & {$\mu \tau$} & {$e \tau$} & {$\mu e$} & {$\mu \tau$} & {$e \tau$} & {$\mu e$} \\
        \midrule
        110.0 & $5.83\times10^{-4}$ & $6.05\times10^{-4}$ & $5.08\times10^{-5}$ & 11.45 & 11.88 & 1.00 \\
        120.0 & $5.86\times10^{-4}$ & $6.11\times10^{-4}$ & $6.38\times10^{-5}$ & 9.07 & 9.45 & 0.99 \\
        125.0 & $5.94\times10^{-4}$ & $6.28\times10^{-4}$ & $7.37\times10^{-5}$ & 7.93 & 8.38 & 0.98 \\
        130.0 & $5.82\times10^{-4}$ & $6.07\times10^{-4}$ & $8.79\times10^{-5}$ & 6.50 & 6.79 & 0.98 \\
        140.0 & $9.80\times10^{-4}$ & $9.90\times10^{-4}$ & $1.53\times10^{-4}$ & 6.34 & 6.40 & 0.99 \\
        150.0 & 0.01 & 0.01 & $1.75\times10^{-3}$ & 6.84 & 6.91 & 1.17 \\
        160.0 & 0.06 & 0.06 & 0.01 & 8.27 & 9.18 & 1.66 \\
        170.0 & 0.13 & 0.16 & 0.03 & 8.96 & 11.40 & 2.10 \\
        180.0 & 0.28 & 0.43 & 0.08 & 9.91 & 15.14 & 2.79 \\
        190.0 & 0.72 & $1.38^{a}$ & 0.26 & 11.63 & 22.14 & 4.17 \\
        200.0 & $2.34^{a}$ & $6.36^{a}$ & $1.25^{a}$ & 14.65 & 39.73 & 7.84 \\
    \end{tabular}
    \end{ruledtabular}
    \caption{Summary of LFV limits. The $\mathcal{B}(H \to \ell\ell')$ is calculated from inclusive $e^+e^- \to \ell^+\ell^- H$ production mode. 
    }
    \footnotesize
    $^{a}$ Branching ratios are unphysical, which are caused by the vanishingly small cross-section with respect to those mass points.
    \label{tab:limits_summary}
\end{table*}

%% file: model.tex
\section{Constraint on Type-III 2HDM}
\label{sec:model}
The type-III 2HDM offers a concrete theoretical framework that could accommodate Higgs LFV decays. The two Higgs doublets are taken to be
\begin{equation}
    H_1 = \begin{pmatrix}G^+\\\frac{v+h_1+iG^0}{\sqrt{2}}\end{pmatrix},\quad
    H_2 = \begin{pmatrix}H^+\\\frac{h_2+iA}{\sqrt{2}}\end{pmatrix},
\end{equation}
where $G^+$ and $G^0$ are the would-be Goldstone bosons, $v=246$~GeV is the electroweak vacuum expectation value, and $H^+$ and $A$ are the charged Higgs and the pseudoscalar boson respectively. The two CP-even neutral scalars $h_1$ and $h_2$ mix to form the mass eigenstates
\begin{equation}
    \begin{pmatrix} h\\H \end{pmatrix} =
    \begin{pmatrix} \cos\alpha &-\sin\alpha\\\sin\alpha&\cos\alpha\end{pmatrix}
    \begin{pmatrix}h_1\\h_2\end{pmatrix}.
\end{equation}
The mixing angle $\alpha$ is constrained by the combined Run 1 and Run 2 measurements of the 125 GeV Higgs boson properties by ATLAS and CMS, which yields $\sin^2\alpha\le0.0225$ at 95\% CL~\cite{ATLAS:2016neq,CMS:2022dwd,ATLAS:2022vkf}.

The LFV interactions are parametrized by the Yukawa coupling of the second doublet. In particular, we take
\begin{equation}
    \mathcal{L}_{LFV} = \sqrt{2}Y_{ij}\bar L_i e_{Rj}H_2.
\end{equation}
Note that the coupling $Y_{ij}$ and $Y_{ji}$, for $i\ne j$, leads to the LFV decays $h (H) \to \ell_i\ell_j$. For simplicity, we will take all components of $Y$ to vanish except for the pair responsible for the LFV decay of interest. In this scenario, the $h$ LFV decay width is given by
\begin{equation}
    \Gamma(h\to\ell\ell') = \frac{\sin^2(\alpha)\, m_h}{8\pi}\left(|Y_{\ell\ell'}|^2 + |Y_{\ell'\ell}|^2\right).
\end{equation}
The LFV decay width for the $H$ can be obtained from the above equation by a replacement $m_h\to m_H$ and $\sin\alpha\to\cos\alpha$.

\begin{figure}[th]
    \centering
    \includegraphics[width=0.5\textwidth]{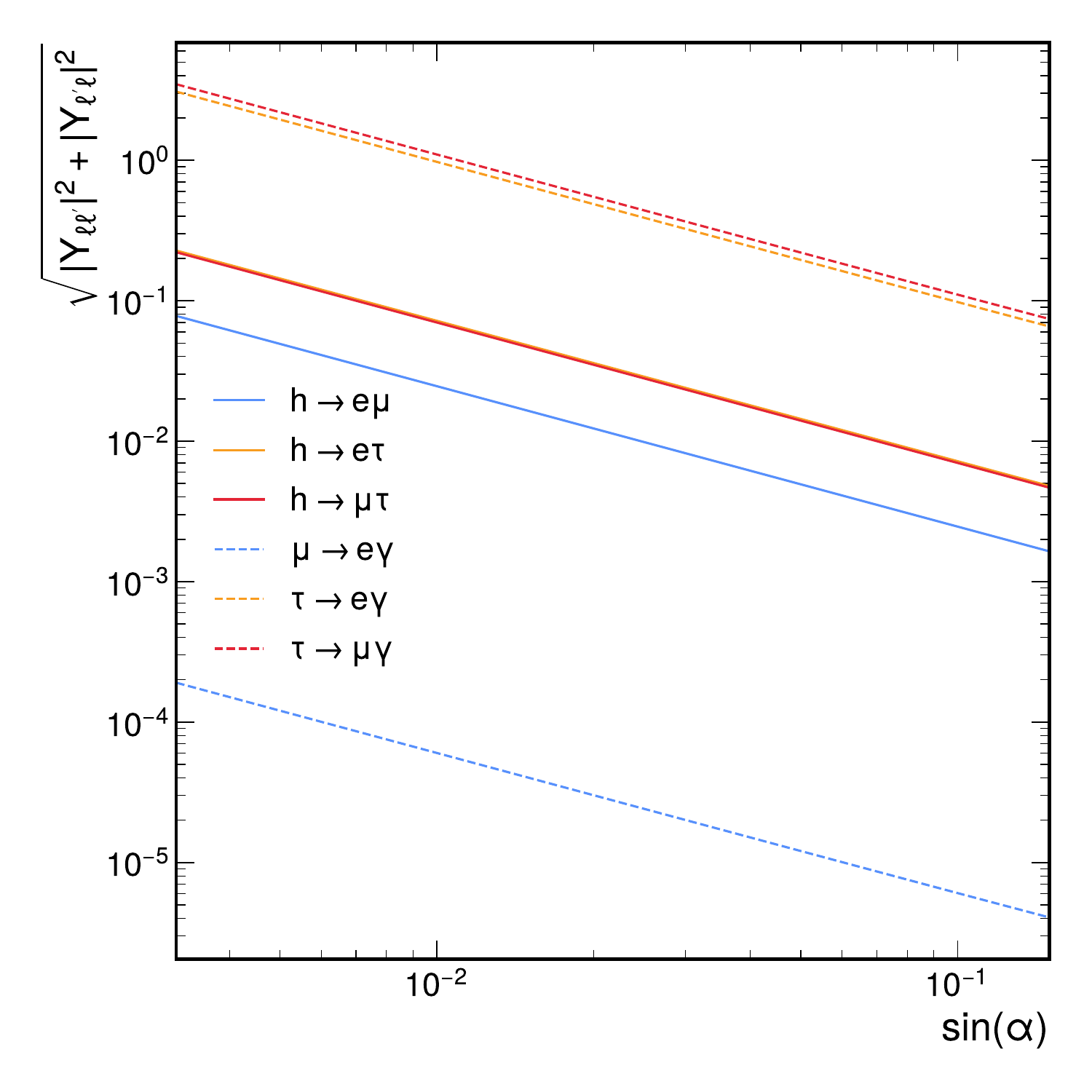}
    \caption{The projected upper limit on the mixing angle, $\sin\alpha$, and the LFV Yukawa couplings, $Y_{\ell\ell'}$ and $Y_{\ell'\ell}$, from the $h\to\ell\ell'$ decay channels (solid). For comparison, the corresponding constraints from the low energy searches $\ell\to\ell'\gamma$ are also shown (dashed). 
    }
    \label{fig:h_sa_Y}
\end{figure}

In the case where the $H$, $A$ and $H^+$ are much heavier than the electroweak scale, our results in the previous section imply 
$\mathcal{B}(h\to e\mu)\le 7.37\times10^{-5}$, $\mathcal{B}(h\to e\tau)\le 6.28\times10^{-4}$ and $\mathcal{B}(h\to \mu\tau)\le 5.94\times10^{-4}$
, respectively.
In the context of the type-III 2HDM, the corresponding branching ratios are given by
\begin{equation}
\begin{aligned}
    \mathcal{B}(h\to\ell\ell') &= \frac{\Gamma(h\to\ell\ell')}{\Gamma(h\to\ell\ell')+\cos^2\alpha\,\Gamma_{h}^{SM}},
    \label{eq:hlfvbr}
\end{aligned}
\end{equation}
with $\Gamma_{h,SM} = 4.1$ MeV being the 125 GeV Higgs boson total decay width. Note that the production of $h$ is also suppressed by $\cos^2(\alpha)$. Thus, the upper 95\% CLs limits on the $h$ LFV branching ratio, $\mathcal{B}(h\to\ell\ell')_{95\%}$, in the previous section translate to
\begin{equation}
\begin{aligned}
    \frac{\cos^2\alpha\,\Gamma(h\to\ell\ell')}{\Gamma(h\to\ell\ell')+\cos^2\alpha\,\Gamma_{h}^{SM}} \le \mathcal{B}(h\to\ell\ell')_{95\%}.
\end{aligned}
\end{equation}
Expanding to lowest order in $\sin^2\alpha$, one gets
\begin{eqnarray}
    (|Y_{\ell\ell'}|^2 + |Y_{\ell'\ell}|^2)\sin^2\alpha {\color{blue}\le} \frac{8\pi\Gamma_{h}^{SM}}{m_h}
    \frac{\mathcal{B}(h\to\ell\ell')_{95\%}}{1-\mathcal{B}(h\to\ell\ell')_{95\%}}
\label{eq:upperlimit}
\end{eqnarray}
Figure~\ref{fig:h_sa_Y} shows the corresponding constraints on the $\sin\alpha$-$\sqrt{|Y_{\ell\ell'}|^2+|Y_{\ell'\ell}|^2}$ plane from the $h\to e\mu$ (solid blue), $h\to e\tau$ (solid orange), and $h\to\mu\tau$ (solid green) decay channels. For comparison, the corresponding constraints from the low energy search for $\ell\to\ell'\gamma$ decays~\cite{MEGII:2025gzr,BaBar:2009hkt,Belle-II:2018jsg,Belle:2021ysv,USBelleIIGroup:2022qro} are shown by dashed lines\footnote{We have used the public Python code provided in Ref.~\cite{Altmannshofer:2025nsl} to compute the $\ell\to\ell'\gamma$ partial decay width.}. Note that in the case of $e-\mu$ LFV, there is also a constraint from $\mu\to e$ conversion in atomic nucleus~\cite{SINDRUMII:2006dvw,Crivellin:2014cta}. Such a constraint is comparable to that of $\mu\to e\gamma$ decay.

\begin{figure}[th]
    \centering
    \includegraphics[width=0.5\textwidth]{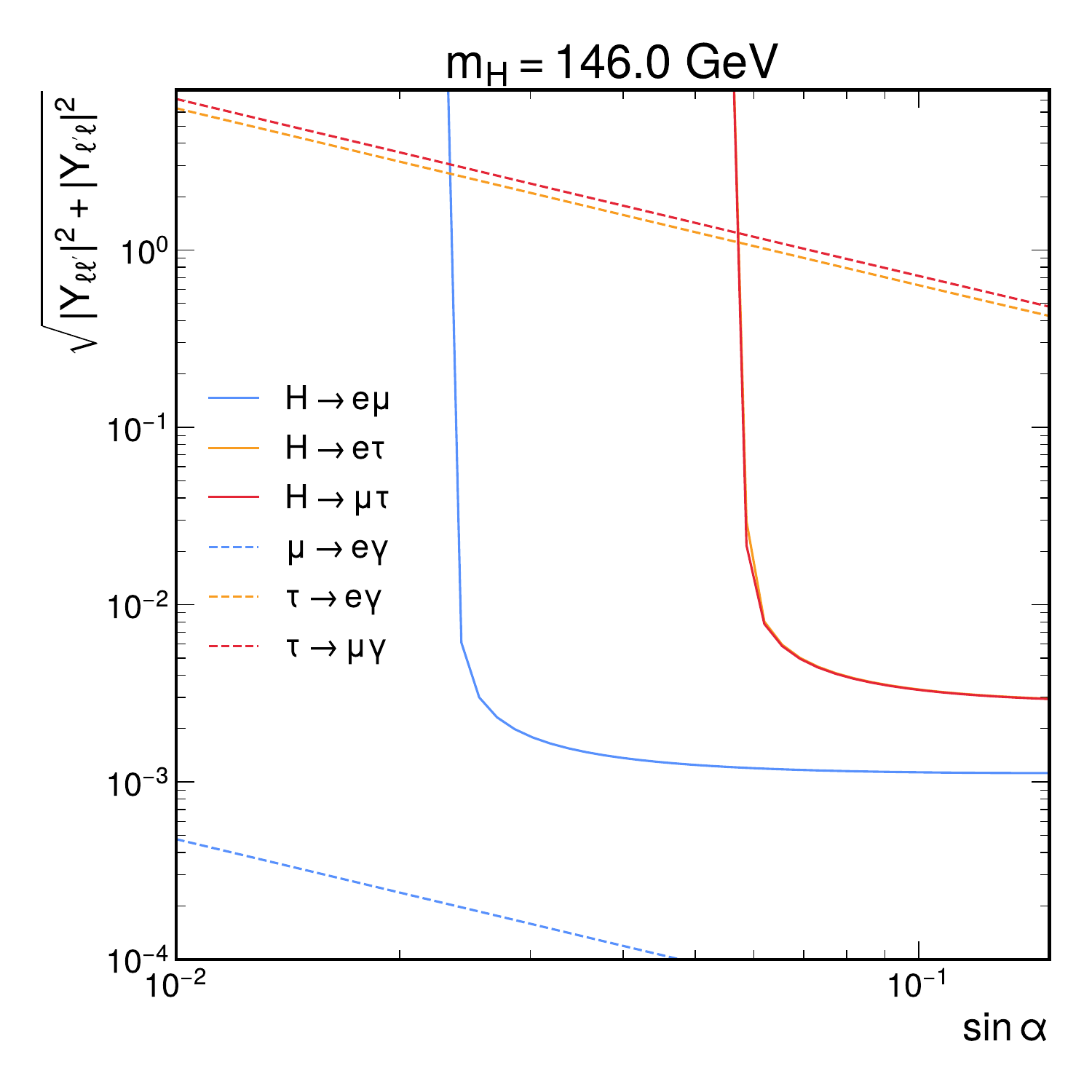}
    \caption{The projected upper limit on the $\sin\alpha-\sqrt{|Y_{\ell\ell'}|^2+|Y_{\ell'\ell}|^2}$ plane from the $H\to\ell\ell'$ decay (solid) and $\ell\to\ell'\gamma$ decay (dashed) for the $e-\mu$ (blue), $e-\tau$ (orange) and $\mu-\tau$ (red) sectors.}
    \label{fig:mH147}
\end{figure}

\begin{figure}[th]
    \centering
    \includegraphics[width=0.5\textwidth]{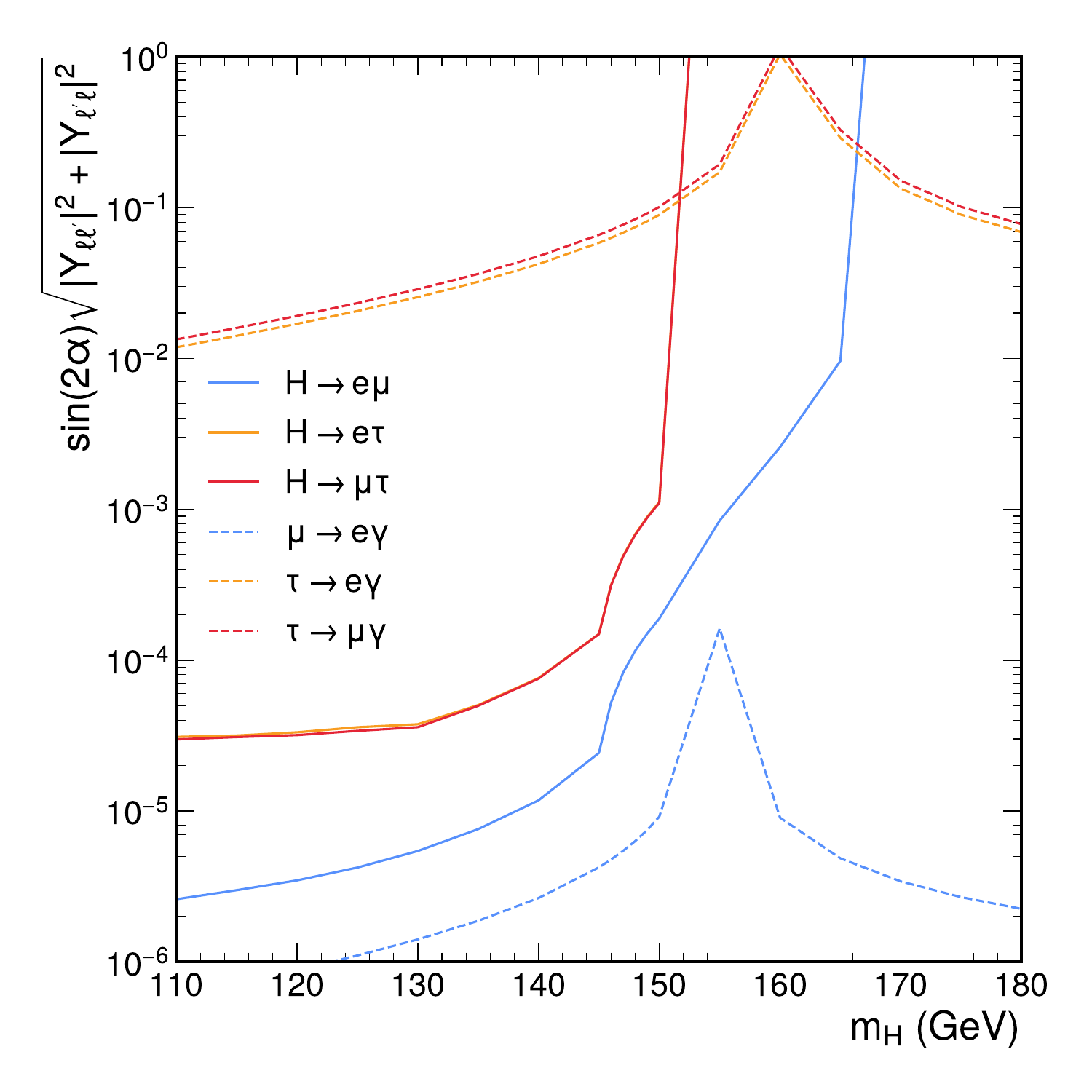}
    \caption{The comparison between the sensitivities of the FCC-ee (solid) and the low energy (dashed) searches for LFV decays in the $e-\mu$ (blue), $e-\tau$ (orange), and $\mu-\tau$ (red) sectors in the mass range 110 GeV $\le m_H\le$ 180 GeV.}
    \label{fig:mH_s2aY}
\end{figure}

Next, we explore the possibility that, in addition to the $h$, the $H$ could also be around the weak scale, while the $H^+$ and $A$ are heavier. In this case, our results from the previous section can be interpreted in terms of the LFV decays of the $H$ in the mass range 110 GeV $\le m_H \le$ 180 GeV. In this scenario, the upper limit on the LFV branching ratios is related to
\begin{equation}
    |Y_{\ell\ell'}|^2+|Y_{\ell'\ell}|^2 \le \tan^2\alpha\frac{8\pi\Gamma_{H}^{SM}}{m_H}\frac{\mathcal{B}(H\to\ell\ell')_{95\%}}{\sin^2\alpha-\mathcal{B}(H\to\ell\ell')_{95\%}},
\end{equation}
where $\Gamma_{H}^{SM}$ is the SM Higgs total decay width at the Higgs mass $m_H$. Note that from the above equation, one can only obtain meaningful constraints when $\mathcal{B}(H\to\ell\ell')_{95\%}\le\sin^2\alpha$. 
In the degenerate case where $m_H\simeq m_h$, one needs to take into account the contribution from both Higgses when extracting the upper limits on the LFV Yukawa couplings.
Figure~\ref{fig:mH147} shows the upper limits on the $\sin\alpha-\sqrt{|Y_{\ell\ell'}|^2+|Y_{\ell'\ell}|^2}$ plane for the case $m_H=146$ GeV. Note that if the lower left corner of the $H\to\ell\ell'$ constraint is below the corresponding  $\ell\to\ell'\gamma$ constraint, the FCC-ee search is more sensitive for this particular value of $m_H$. To compare the sensitivity of the FCC-ee and the low energy searches across the $m_H$ mass range, we follow Ref.~\cite{Primulando:2023ugc} in projecting the constraints onto the $\sin(2\alpha)\sqrt{|Y_{\ell\ell'}|^2+|Y_{\ell'\ell}|^2}$ line. For the FCC-ee constraints, we take the minimum value of such projections. The comparison between the two types of searches for the LFV couplings is shown in Figure~\ref{fig:mH_s2aY}.

%% file: summary.tex
\section{Summary and Conclusions}
\label{sec:summary}

In this analysis, we have explored the prospects of detecting the LFV decays of $H$ in the $\htomue$, $\htomutau$, and $\htoetau$ channels at the FCC-ee with an integrated luminosity of 5 ab$^{-1}$ at $\sqrt{s} = 240$ GeV. Our analysis focuses on clean 4-lepton final states, which require the $\tau$ to decay leptonically. For the three LFV channels, both $ZH$ and VBF production modes are considered together via the inclusive process $e^+ e^- \to \ell^+ \ell^- H$ to account for the interference effect, as discussed in Section~\ref{sec:mc_simulation}
. This allows us to probe $H$ masses up to 190 GeV for the $\htomue$ and $\htomutau$ channels, and up to 180 GeV for $\htoetau$ mode. For the Higgs mass in the range $110-200$ GeV we have derived the expected 95\% CL upper limits on the branching fractions of the LFV Higgs decays, shown in Table~\ref{tab:limits_summary}. At $m_H = 125$ GeV, the combined limits are found to be 
$\mathcal{B}(\htomutau) < 5.94 \times 10^{-4}$, $\mathcal{B}(\htoetau) < 6.28 \times 10^{-4}$, and $\mathcal{B}(\htomue) < 7.37 \times 10^{-5}$.

We have also discussed the constraints on BSM models, particularly the type-III 2HDM, given the expected limits on the LFV Higgs decays obtained in this analysis. As shown in Figure~\ref{fig:h_sa_Y}, the expected limits of $H$ LFV decays in $e \tau$ and $\mu \tau$ sectors provide better constraints on the mixing angle $\sin \alpha$ and the LFV Yukawa couplings compared to the limit derived from low energy $\tau\to\ell\gamma$ searches. 

The findings presented in this paper indicate that the FCC-ee detector, proposed for construction and operation, will provide new insights into the study of BSM physics. Still, there are a number of assumptions that have to be made due to the lack of data from the yet-to-be-constructed collider itself, especially the assumptions about systematic uncertainties. Future studies on the expected systematic uncertainties that would apply to future colliders and detectors will provide additional insights and a better estimation of the accuracy for these searches.

%% file: bibtex.bib
@article{PhysRevD.104.032013,
  title = "{Search for lepton-flavor violating decays of the Higgs boson in the $\mu\tau$ and $e\ensuremath{\tau}$ final states in proton-proton collisions at $\sqrt{s}=13 \mathrm{ TeV}$}",
  author = {Sirunyan, A. M. and others},
  collaboration = {CMS Collaboration},
  journal = {Phys. Rev. D},
  volume = {104},
  issue = {3},
  pages = {032013},
  numpages = {31},
  year = {2021},
  month = {August},
  publisher = {American Physical Society},
  doi = {10.1103/PhysRevD.104.032013},
  url = {https://link.aps.org/doi/10.1103/PhysRevD.104.032013}
}

@article{Branco:2011iw,
    author = "Branco, G. C. and Ferreira, P. M. and Lavoura, L. and Rebelo, M. N. and Sher, Marc and Silva, Joao P.",
    title = "{Theory and phenomenology of two-Higgs-doublet models}",
    eprint = "1106.0034",
    archivePrefix = "arXiv",
    primaryClass = "hep-ph",
    doi = "10.1016/j.physrep.2012.02.002",
    journal = "Phys. Rept.",
    volume = "516",
    pages = "1--102",
    year = "2012"
}

@article{Primulando:2023ugc,
    author = "Primulando, R. and Julio, J. and Srimanobhas, N. and Uttayarat, P.",
    title = "{A new Higgs boson with electron-muon flavor-violating couplings}",
    eprint = "2304.13757",
    archivePrefix = "arXiv",
    primaryClass = "hep-ph",
    doi = "10.1016/j.physletb.2023.138129",
    journal = "Phys. Lett. B",
    volume = "845",
    pages = "138129",
    year = "2023"
}

@article{Belle:2021ysv,
    author = "Abdesselam, A. and others",
    collaboration = "Belle",
    title = "{Search for lepton-flavor-violating tau-lepton decays to $\ell\gamma$ at Belle}",
    eprint = "2103.12994",
    archivePrefix = "arXiv",
    primaryClass = "hep-ex",
    doi = "10.1007/JHEP10(2021)019",
    journal = "J. High Energy Phys.",
    volume = "10",
    pages = "19",
    year = "2021"
}

@article{BaBar:2009hkt,
    author = "Aubert, Bernard and others",
    collaboration = "BaBar",
    title = "{Searches for Lepton Flavor Violation in the Decays $\tau^\pm \to e^\pm \gamma$ and $\tau^\pm \to \mu^\pm \gamma$}",
    eprint = "0908.2381",
    archivePrefix = "arXiv",
    primaryClass = "hep-ex",
    reportNumber = "SLAC-PUB-13753, BABAR-PUB-09-026",
    doi = "10.1103/PhysRevLett.104.021802",
    journal = "Phys. Rev. Lett.",
    volume = "104",
    pages = "021802",
    year = "2010"
}

@article{MEGII:2025gzr,
    author = "Afanaciev, K. and others",
    collaboration = "MEG II",
    title = "{New limit on the ${\mu^+ \rightarrow e^+ \gamma }$ decay with the MEG II experiment}",
    eprint = "2504.15711",
    archivePrefix = "arXiv",
    primaryClass = "hep-ex",
    doi = "10.1140/epjc/s10052-025-14906-3",
    journal = "Eur. Phys. J. C",
    volume = "85",
    number = "10",
    pages = "1177",
    year = "2025",
    note = "[Erratum: Eur.Phys.J.C 85, 1317 (2025)]"
}

@article{Altmannshofer:2025nsl,
    author = "Altmannshofer, Wolfgang and Assi, Beno{\^\i}t and Brod, Joachim and Hamer, Nick and Julio, J. and Uttayarat, Patipan and Volkov, Daniil",
    title = "{Electron EDM and {\ensuremath{\Gamma}}({\ensuremath{\mu}} {\textrightarrow} e{\ensuremath{\gamma}}) in the 2HDM}",
    eprint = "2410.17313",
    archivePrefix = "arXiv",
    primaryClass = "hep-ph",
    doi = "10.1007/JHEP06(2025)156",
    journal = "J. High Energy Phys.",
    volume = "06",
    pages = "156",
    year = "2025"
}

@article{ATLAS:2016neq,
    author = "Aad, Georges and others",
    collaboration = "ATLAS, CMS",
    title = "{Measurements of the Higgs boson production and decay rates and constraints on its couplings from a combined ATLAS and CMS analysis of the LHC pp collision data at $ \sqrt{s}=7 $ and 8 TeV}",
    eprint = "1606.02266",
    archivePrefix = "arXiv",
    primaryClass = "hep-ex",
    reportNumber = "CERN-EP-2016-100, ATLAS-HIGG-2015-07, CMS-HIG-15-002",
    doi = "10.1007/JHEP08(2016)045",
    journal = "J. High Energy Phys.",
    volume = "08",
    pages = "045",
    year = "2016"
}

@article{CMS:2022dwd,
    author = "Tumasyan, Armen and others",
    collaboration = "CMS",
    title = "{A portrait of the Higgs boson by the CMS experiment ten years after the discovery.}",
    eprint = "2207.00043",
    archivePrefix = "arXiv",
    primaryClass = "hep-ex",
    reportNumber = "CMS-HIG-22-001, CERN-EP-2022-039",
    doi = "10.1038/s41586-022-04892-x",
    journal = "Nature",
    volume = "607",
    number = "7917",
    pages = "60--68",
    year = "2022",
    note = "[Erratum: Nature 623, (2023)]"
}

@article{ATLAS:2022vkf,
    author = "Aad, Georges and others",
    collaboration = "ATLAS",
    title = "{A detailed map of Higgs boson interactions by the ATLAS experiment ten years after the discovery}",
    eprint = "2207.00092",
    archivePrefix = "arXiv",
    primaryClass = "hep-ex",
    reportNumber = "CERN-EP-2022-057",
    doi = "10.1038/s41586-022-04893-w",
    journal = "Nature",
    volume = "607",
    number = "7917",
    pages = "52--59",
    year = "2022",
    note = "[Erratum: Nature 612, E24 (2022)]"
}

@article{LHCb:2018ukt,
    author = "Aaij, Roel and others",
    collaboration = "LHCb",
    title = "{Search for lepton-flavour-violating decays of Higgs-like bosons}",
    eprint = "1808.07135",
    archivePrefix = "arXiv",
    primaryClass = "hep-ex",
    reportNumber = "CERN-EP-2018-210, LHCb-PAPER-2018-030",
    doi = "10.1140/epjc/s10052-018-6386-8",
    journal = "Eur. Phys. J. C",
    volume = "78",
    number = "12",
    pages = "1008",
    year = "2018"
}

@article{CMS:2019pex,
    author = "Sirunyan, Albert M and others",
    collaboration = "CMS",
    title = "{Search for lepton flavour violating decays of a neutral heavy Higgs boson to $\mu\tau$ and e$\tau$ in proton-proton collisions at $\sqrt{s}=$ 13 TeV}",
    eprint = "1911.10267",
    archivePrefix = "arXiv",
    primaryClass = "hep-ex",
    reportNumber = "CMS-HIG-18-017, CERN-EP-2019-170",
    doi = "10.1007/JHEP03(2020)103",
    journal = "J. High Energy Phys.",
    volume = "03",
    pages = "103",
    year = "2020"
}

@article{ATLAS:2019old,
    author = "Aad, Georges and others",
    collaboration = "ATLAS",
    title = "{Search for the Higgs boson decays $H \to ee$ and $H \to e\mu$ in $pp$ collisions at $\sqrt{s} = 13$ TeV with the ATLAS detector}",
    eprint = "1909.10235",
    archivePrefix = "arXiv",
    primaryClass = "hep-ex",
    reportNumber = "CERN-EP-2019-184",
    doi = "10.1016/j.physletb.2019.135148",
    journal = "Phys. Lett. B",
    volume = "801",
    pages = "135148",
    year = "2020"
}

@article{Casagrande:2008hr,
    author = "Casagrande, S. and Goertz, F. and Haisch, U. and Neubert, M. and Pfoh, T.",
    title = "{Flavor Physics in the Randall-Sundrum Model: I. Theoretical Setup and Electroweak Precision Tests}",
    eprint = "0807.4937",
    archivePrefix = "arXiv",
    primaryClass = "hep-ph",
    reportNumber = "MZ-TH-08-18",
    doi = "10.1088/1126-6708/2008/10/094",
    journal = "JHEP",
    volume = "10",
    pages = "094",
    year = "2008"
}

@article{Han:2000jz,
    author = "Han, Tao and Marfatia, Danny",
    title = "{h ---{\ensuremath{>}} mu tau at hadron colliders}",
    eprint = "hep-ph/0008141",
    archivePrefix = "arXiv",
    reportNumber = "MADPH-00-1188",
    doi = "10.1103/PhysRevLett.86.1442",
    journal = "Phys. Rev. Lett.",
    volume = "86",
    pages = "1442--1445",
    year = "2001"
}

@article{Arhrib:2012ax,
    author = "Arhrib, Abdesslam and Cheng, Yifan and Kong, Otto C. W.",
    title = "{Comprehensive analysis on lepton flavor violating Higgs boson to $\mu^\mp \tau^\pm$ decay in supersymmetry without $R$ parity}",
    eprint = "1210.8241",
    archivePrefix = "arXiv",
    primaryClass = "hep-ph",
    reportNumber = "NCU-HEP-K055",
    doi = "10.1103/PhysRevD.87.015025",
    journal = "Phys. Rev. D",
    volume = "87",
    number = "1",
    pages = "015025",
    year = "2013"
}

@article{Albrecht:2009xr,
    author = "Albrecht, Michaela E. and Blanke, Monika and Buras, Andrzej J. and Duling, Bjorn and Gemmler, Katrin",
    title = "{Electroweak and Flavour Structure of a Warped Extra Dimension with Custodial Protection}",
    eprint = "0903.2415",
    archivePrefix = "arXiv",
    primaryClass = "hep-ph",
    reportNumber = "TUM-TEP-711-09, MPP-2009-17, TUM-HEP-711-09",
    doi = "10.1088/1126-6708/2009/09/064",
    journal = "JHEP",
    volume = "09",
    pages = "064",
    year = "2009"
}

@article{Altmannshofer:2016zrn,
    author = "Altmannshofer, Wolfgang and Eby, Joshua and Gori, Stefania and Lotito, Matteo and Martone, Mario and Tuckler, Douglas",
    title = "{Collider Signatures of Flavorful Higgs Bosons}",
    eprint = "1610.02398",
    archivePrefix = "arXiv",
    primaryClass = "hep-ph",
    reportNumber = "FERMILAB-PUB-16-499-PPD",
    doi = "10.1103/PhysRevD.94.115032",
    journal = "Phys. Rev. D",
    volume = "94",
    number = "11",
    pages = "115032",
    year = "2016"
}

@article{Perez:2008ee,
    author = "Perez, Gilad and Randall, Lisa",
    title = "{Natural Neutrino Masses and Mixings from Warped Geometry}",
    eprint = "0805.4652",
    archivePrefix = "arXiv",
    primaryClass = "hep-ph",
    doi = "10.1088/1126-6708/2009/01/077",
    journal = "JHEP",
    volume = "01",
    pages = "077",
    year = "2009"
}

@article{CMS:2023pte,
    author = "Hayrapetyan, Aram and others",
    collaboration = "CMS",
    title = "{Search for the lepton-flavor violating decay of the Higgs boson and additional Higgs bosons in the e$\mu$ final state in proton-proton collisions at $\sqrt{s}$ = 13 TeV}",
    eprint = "2305.18106",
    archivePrefix = "arXiv",
    primaryClass = "hep-ex",
    reportNumber = "CMS-HIG-22-002, CERN-EP-2023-061",
    doi = "10.1103/PhysRevD.108.072004",
    journal = "Phys. Rev. D",
    volume = "108",
    number = "7",
    pages = "072004",
    year = "2023"
}

@article{JHEP.07.2023.166,
   title="{Searches for lepton-flavour-violating decays of the Higgs boson into $e\tau$ and $\mu\tau$ in $\sqrt{s}$ = 13 TeV $pp$ collisions with the ATLAS detector}",
   volume={2023},
   ISSN={1029-8479},
   url={http://dx.doi.org/10.1007/JHEP07(2023)166},
   DOI={10.1007/jhep07(2023)166},
   number={7},
   journal={J. High Energy Phys.},
   publisher={Springer Science and Business Media LLC},
   author={Aad, G. and Abbott, B. and Abbott, D. C. and Abeling, K. and Abidi, S. H.},
   year={2023},
   month=jul }

@article{CMS:2012qbp,
    author = {S. Chatrchyan and others},
    collaboration = "CMS",
    title = "{Observation of a New Boson at a Mass of 125 GeV with the CMS Experiment at the LHC}",
    eprint = "1207.7235",
    archivePrefix = "arXiv",
    primaryClass = "hep-ex",
    reportNumber = "CMS-HIG-12-028, CERN-PH-EP-2012-220",
    doi = "10.1016/j.physletb.2012.08.021",
    journal = "Phys. Lett. B",
    volume = "716",
    pages = "30--61",
    year = "2012"
}

@article{ATLAS:2012yve,
    author = {Aad, G. and others},
    collaboration = "ATLAS",
    title = "{Observation of a new particle in the search for the Standard Model Higgs boson with the ATLAS detector at the LHC}",
    eprint = "1207.7214",
    archivePrefix = "arXiv",
    primaryClass = "hep-ex",
    reportNumber = "CERN-PH-EP-2012-218",
    doi = "10.1016/j.physletb.2012.08.020",
    journal = "Phys. Lett. B",
    volume = "716",
    pages = "1--29",
    year = "2012"
}

@article{MadGraph,
    author = "Alwall, J. and others",
    title = "{The automated computation of tree-level and next-to-leading order differential cross sections, and their matching to parton shower simulations}",
    eprint = "1405.0301",
    archivePrefix = "arXiv",
    primaryClass = "hep-ph",
    reportNumber = "CERN-PH-TH-2014-064, CP3-14-18, LPN14-066, MCNET-14-09, ZU-TH-14-14",
    doi = "10.1007/JHEP07(2014)079",
    journal = "J. High Energy Phys.",
    volume = "07",
    pages = "079",
    year = "2014"
}

@article{Pythia,
    author = {Sj\"ostrand, T. and others},
    title = "{An introduction to PYTHIA 8.2}",
    eprint = "1410.3012",
    archivePrefix = "arXiv",
    primaryClass = "hep-ph",
    reportNumber = "LU-TP-14-36, MCNET-14-22, CERN-PH-TH-2014-190, FERMILAB-PUB-14-316-CD, DESY-14-178, SLAC-PUB-16122",
    doi = "10.1016/j.cpc.2015.01.024",
    journal = "Comput. Phys. Commun.",
    volume = "191",
    pages = "159--177",
    year = "2015"
}

@misc{Delphes,
    author = "Ovyn, S. and Rouby, X. and Lemaitre, V.",
    title = "{DELPHES, a framework for fast simulation of a generic collider experiment}",
    eprint = "0903.2225",
    archivePrefix = "arXiv",
    primaryClass = "hep-ph",
    reportNumber = "CP3-09-01",
    month = "March",
    year = "2009"
}

@article{Belle-II:2018jsg,
    author = "Altmannshofer, W. and others",
    editor = "Kou, E. and Urquijo, P.",
    collaboration = "Belle-II",
    title = "{The Belle II Physics Book}",
    eprint = "1808.10567",
    archivePrefix = "arXiv",
    primaryClass = "hep-ex",
    reportNumber = "KEK Preprint 2018-27, BELLE2-PUB-PH-2018-001, FERMILAB-PUB-18-398-T, JLAB-THY-18-2780, INT-PUB-18-047, UWThPh 2018-26",
    doi = "10.1093/ptep/ptz106",
    journal = "PTEP",
    volume = "2019",
    number = "12",
    pages = "123C01",
    year = "2019",
    note = "[Erratum: PTEP 2020, 029201 (2020)]"
}

@article{SINDRUMII:2006dvw,
    author = "Bertl, Wilhelm H. and others",
    collaboration = "SINDRUM II",
    title = "{A Search for muon to electron conversion in muonic gold}",
    doi = "10.1140/epjc/s2006-02582-x",
    journal = "Eur. Phys. J. C",
    volume = "47",
    pages = "337--346",
    year = "2006"
}

@article{Crivellin:2014cta,
    author = "Crivellin, Andreas and Hoferichter, Martin and Procura, Massimiliano",
    title = "{Improved predictions for $\mu\to e$ conversion in nuclei and Higgs-induced lepton flavor violation}",
    eprint = "1404.7134",
    archivePrefix = "arXiv",
    primaryClass = "hep-ph",
    reportNumber = "CERN-PH-TH-2014-068",
    doi = "10.1103/PhysRevD.89.093024",
    journal = "Phys. Rev. D",
    volume = "89",
    pages = "093024",
    year = "2014"
}

@inproceedings{USBelleIIGroup:2022qro,
    author = "Asner, D. M. and others",
    collaboration = "US Belle II Group, Belle II/SuperKEKB e- Polarization Upgrade Working Group",
    title = "{Snowmass 2021 White Paper on Upgrading SuperKEKB with a Polarized Electron Beam: Discovery Potential and Proposed Implementation}",
    booktitle = "{Snowmass 2021}",
    eprint = "2205.12847",
    archivePrefix = "arXiv",
    primaryClass = "physics.acc-ph",
    month = "5",
    year = "2022"
}

@article{Arganda:2019gnv,
    author = "Arganda, Ernesto and Marcano, Xabier and Mileo, Nicol{\'a}s I. and Morales, Roberto A. and Szynkman, Alejandro",
    title = "{Model-independent search strategy for the lepton-flavor-violating heavy Higgs boson decay to $\tau\mu$ at the LHC}",
    eprint = "1906.08282",
    archivePrefix = "arXiv",
    primaryClass = "hep-ph",
    reportNumber = "IFT-UAM/CSIC-19-78, FTUAM-19-12, LPT-Orsay-19-14",
    doi = "10.1140/epjc/s10052-019-7249-7",
    journal = "Eur. Phys. J. C",
    volume = "79",
    number = "9",
    pages = "738",
    year = "2019"
}

@article{Afik:2023vyl,
    author = "Afik, Yoav and Bhupal Dev, P. S. and Thapa, Anil",
    title = "{Hints of a new leptophilic Higgs sector?}",
    eprint = "2305.19314",
    archivePrefix = "arXiv",
    primaryClass = "hep-ph",
    doi = "10.1103/PhysRevD.109.015003",
    journal = "Phys. Rev. D",
    volume = "109",
    number = "1",
    pages = "015003",
    year = "2024"
}

@article{Urena:2021xtw,
    author = "Ure{\~n}a, Marco Antonio Arroyo and Gaitan-Lozano, Ricardo and de Oca Yemha, Jos{\'e} Halim Montes and V{\'e}lez, Ricardo S{\'a}nchez",
    title = "{Lepton flavor violating h{\textrightarrow}{\ensuremath{\tau}}{\ensuremath{\mu}} decay induced by leptoquarks}",
    doi = "10.31349/RevMexFis.67.040801",
    journal = "Rev. Mex. Fis.",
    volume = "67",
    number = "4",
    pages = "040801",
    year = "2021"
}

@article{Barman:2022iwj,
    author = "Barman, Rahool Kumar and Dev, P. S. Bhupal and Thapa, Anil",
    title = "{Constraining lepton flavor violating Higgs couplings at the HL-LHC in the vector boson fusion channel}",
    eprint = "2210.16287",
    archivePrefix = "arXiv",
    primaryClass = "hep-ph",
    doi = "10.1103/PhysRevD.107.075018",
    journal = "Phys. Rev. D",
    volume = "107",
    number = "7",
    pages = "075018",
    year = "2023"
}

@article{Primulando:2022vip,
    author = "Primulando, R. and Julio, J. and Uttayarat, P.",
    title = "{Minimal Zee model for lepton g-2 and W-mass shifts}",
    eprint = "2211.16021",
    archivePrefix = "arXiv",
    primaryClass = "hep-ph",
    doi = "10.1103/PhysRevD.107.055034",
    journal = "Phys. Rev. D",
    volume = "107",
    number = "5",
    pages = "055034",
    year = "2023"
}

@article{Primulando:2022lpj,
    author = "Primulando, R. and Julio, J. and Uttayarat, P.",
    title = "{Zee model with quasidegenerate neutrino masses and where to find it}",
    eprint = "2201.01960",
    archivePrefix = "arXiv",
    primaryClass = "hep-ph",
    doi = "10.1140/epjc/s10052-022-10189-0",
    journal = "Eur. Phys. J. C",
    volume = "82",
    number = "3",
    pages = "253",
    year = "2022"
}

@article{Primulando:2016eod,
    author = "Primulando, Reinard and Uttayarat, Patipan",
    title = "{Probing Lepton Flavor Violation at the 13 TeV LHC}",
    eprint = "1612.01644",
    archivePrefix = "arXiv",
    primaryClass = "hep-ph",
    doi = "10.1007/JHEP05(2017)055",
    journal = "JHEP",
    volume = "05",
    pages = "055",
    year = "2017"
}

@article{Primulando:2019ydt,
    author = "Primulando, R. and Julio, J. and Uttayarat, P.",
    title = "{Collider Constraints on Lepton Flavor Violation in the 2HDM}",
    eprint = "1912.08533",
    archivePrefix = "arXiv",
    primaryClass = "hep-ph",
    doi = "10.1103/PhysRevD.101.055021",
    journal = "Phys. Rev. D",
    volume = "101",
    number = "5",
    pages = "055021",
    year = "2020"
}

@article{
    CMS-Higgs-Combine,
    author = {Hayrapetyan, A. and others},
    title = "The {CMS} statistical analysis and combination tool: {\textsc{Combine}}",
    eprint = "2404.06614",
    archivePrefix = "arXiv",
    primaryClass = "physics.data-an",
    reportNumber = "CMS-CAT-23-001, CERN-EP-2024-078",
    year = "2024",
    journal = "Comput. Softw. Big Sci.",
    doi = "10.1007/s41781-024-00121-4",
    volume = "8",
    pages = "19"
}

@misc{IDEA,
      title="{The IDEA detector concept for FCC-ee}", 
      author="{The IDEA Study Group}",
      year={2025},
      eprint={2502.21223},
      archivePrefix={arXiv},
      primaryClass={physics.ins-det},
      url={https://arxiv.org/abs/2502.21223}, 
}

@article{FCCeePlan,
    author = {Benedikt, Michael and Blondel, Alain and Janot, Patrick and Mangano, Michelangelo and Zimmermann, Frank},
    year = {2020},
    month = {04},
    pages = {402-407},
    title = "{Future Circular Colliders succeeding the LHC}",
    volume = {16},
    journal = {Nature Physics},
    doi = {10.1038/s41567-020-0856-2}
}

@article{Eur.Phys.J.C76.2016.6,
   title="{Measurements of the Higgs boson production and decay rates and coupling strengths using $pp$ collision data at $\sqrt{s}=7$ and 8 TeV in the ATLAS experiment}",
   volume={76},
   ISSN={1434-6052},
   url={https://dx.doi.org/10.1140/epjc/s10052-015-3769-y},
   DOI={10.1140/epjc/s10052-015-3769-y},
   number={1},
   journal={Eur. Phys. J. C},
   publisher={Springer Science and Business Media LLC},
   author={Aad, G. and others},
   year={2016},
   month=jan 
}

@article{Eur.Phys.J.C.75.2015.212,
   title="{Precise determination of the mass of the Higgs boson and tests of compatibility of its couplings with the standard model predictions using proton collisions at 7 and 8 $\,\text {TeV}$}",
   volume={75},
   ISSN={1434-6052},
   url={https://dx.doi.org/10.1140/epjc/s10052-015-3351-7},
   DOI={10.1140/epjc/s10052-015-3351-7},
   number={5},
   journal={Eur. Phys. J. C},
   publisher={Springer Science and Business Media LLC},
   author={Khachatryan, V. and others},
   year={2015},
   month=may 
}

@article{Phys.Rev.D111.2025.092014,
   title={Measurement of the Higgs boson mass and width using the four-lepton final state in proton-proton collisions at $\sqrt{s}=13$},
   volume={111},
   ISSN={2470-0029},
   url={https://dx.doi.org/10.1103/PhysRevD.111.092014},
   DOI={10.1103/physrevd.111.092014},
   number={9},
   journal={Phys. Rev. D},
   publisher={American Physical Society (APS)},
   author={Hayrapetyan, A. and others},
   year={2025},
   month=may }

@article{Diaz_Cruz_2000,
   title={Lepton flavor violating decays of Higgs bosons beyond the standard model},
   volume={62},
   ISSN={1089-4918},
   url={http://dx.doi.org/10.1103/PhysRevD.62.116005},
   DOI={10.1103/physrevd.62.116005},
   number={11},
   journal={Phys. Rev. D},
   publisher={American Physical Society (APS)},
   author={Diaz-Cruz, J. Lorenzo and Toscano, J. J.},
   year={2000},
   month=nov }

@article{Elagin_2011,
   title={A new mass reconstruction technique for resonances decaying to $\tau\tau$},
   volume={654},
   ISSN={0168-9002},
   url={http://dx.doi.org/10.1016/j.nima.2011.07.009},
   DOI={10.1016/j.nima.2011.07.009},
   number={1},
   journal={Nuclear Instruments and Methods in Physics Research Section A: Accelerators, Spectrometers, Detectors and Associated Equipment},
   publisher={Elsevier BV},
   author={Elagin, A. and Murat, P. and Pranko, A. and Safonov, A.},
   year={2011},
   month=oct, pages={481–489} 
   }

@article{barchetta2021trackingvertexdetectorsfccee,
      title={Tracking and Vertex detectors at FCC-ee}, 
      author={Nicola Barchetta and Paula Collins and Petra Riedler},
      year={2021},
      eprint={2112.13019},
      archivePrefix={arXiv},
      primaryClass={physics.ins-det},
      url={https://arxiv.org/abs/2112.13019}
}

@article{Qin_2018,
   title={Charged lepton flavor violating Higgs decays at future $e^+e^-$ colliders},
   volume={78},
   ISSN={1434-6052},
   url={http://dx.doi.org/10.1140/epjc/s10052-018-6298-7},
   DOI={10.1140/epjc/s10052-018-6298-7},
   number={10},
   journal={Eur. Phys. J. C},
   publisher={Springer Science and Business Media LLC},
   author={Qin, Qin and Li, Qiang and L\"u, Cai-Dian and Yu, Fu-Sheng and Zhou, Si-Hong},
   year={2018},
   month=oct }

@misc{Frixione:2021zdp,
    author = "Frixione, Stefano and Mattelaer, Olivier and Zaro, Marco and Zhao, Xiaoran",
    title = "{Lepton collisions in MadGraph5{\_}aMC@NLO}",
    eprint = "2108.10261",
    archivePrefix = "arXiv",
    primaryClass = "hep-ph",
    reportNumber = "MCNET-21-13,CP3-21-50",
    month = "8",
    year = "2021"
}

@article{Arana-Catania:2013xma,
    author = "Arana-Catania, M. and Arganda, E. and Herrero, M. J.",
    title = "{Non-decoupling SUSY in LFV Higgs decays: a window to new physics at the LHC}",
    eprint = "1304.3371",
    archivePrefix = "arXiv",
    primaryClass = "hep-ph",
    reportNumber = "IFT-UAM-CSIC-13-024",
    doi = "10.1007/JHEP09(2013)160",
    journal = "JHEP",
    volume = "09",
    pages = "160",
    year = "2013",
    note = "[Erratum: JHEP 10, 192 (2015)]"
}

@article{Arganda:2015naa,
    author = "Arganda, E. and Herrero, M. J. and Marcano, X. and Weiland, C.",
    title = "{Enhancement of the lepton flavor violating Higgs boson decay rates from SUSY loops in the inverse seesaw model}",
    eprint = "1508.04623",
    archivePrefix = "arXiv",
    primaryClass = "hep-ph",
    reportNumber = "IFT-UAM-CSIC-15-085, FTUAM-15-25, SU-HET-07-2015, IPPP-15-67, DCPT-15-134",
    doi = "10.1103/PhysRevD.93.055010",
    journal = "Phys. Rev. D",
    volume = "93",
    number = "5",
    pages = "055010",
    year = "2016"
}

@article{Arganda:2015uca,
    author = "Arganda, E. and Herrero, M. J. and Morales, R. and Szynkman, A.",
    title = "{Analysis of the h, H, A {\textrightarrow} {\ensuremath{\tau}}{\ensuremath{\mu}} decays induced from SUSY loops within the Mass Insertion Approximation}",
    eprint = "1510.04685",
    archivePrefix = "arXiv",
    primaryClass = "hep-ph",
    reportNumber = "IFT-UAM-CSIC-15-108, FTUAM-15-32",
    doi = "10.1007/JHEP03(2016)055",
    journal = "JHEP",
    volume = "03",
    pages = "055",
    year = "2016"
}

@article{Arganda:2004bz,
    author = "Arganda, Ernesto and Curiel, Ana M. and Herrero, Maria J. and Temes, David",
    title = "{Lepton flavor violating Higgs boson decays from massive seesaw neutrinos}",
    eprint = "hep-ph/0407302",
    archivePrefix = "arXiv",
    reportNumber = "FTUAM-04-16, IFT-UAM-CSIC-04-41, LAPTH-1057-04",
    doi = "10.1103/PhysRevD.71.035011",
    journal = "Phys. Rev. D",
    volume = "71",
    pages = "035011",
    year = "2005"
}

@article{Arganda:2014dta,
    author = "Arganda, E. and Herrero, M. J. and Marcano, X. and Weiland, C.",
    title = "{Imprints of massive inverse seesaw model neutrinos in lepton flavor violating Higgs boson decays}",
    eprint = "1405.4300",
    archivePrefix = "arXiv",
    primaryClass = "hep-ph",
    reportNumber = "IFT-UAM-CSIC-14-018, FTUAM-14-016",
    doi = "10.1103/PhysRevD.91.015001",
    journal = "Phys. Rev. D",
    volume = "91",
    number = "1",
    pages = "015001",
    year = "2015"
}

@article{Arganda:2016zvc,
    author = "Arganda, E. and Herrero, M. J. and Marcano, X. and Morales, R. and Szynkman, A.",
    title = "{Effective lepton flavor violating H{\ensuremath{\ell}}i{\ensuremath{\ell}}j vertex from right-handed neutrinos within the mass insertion approximation}",
    eprint = "1612.09290",
    archivePrefix = "arXiv",
    primaryClass = "hep-ph",
    reportNumber = "IFT-UAM-CSIC-16-137, FTUAM-16-46",
    doi = "10.1103/PhysRevD.95.095029",
    journal = "Phys. Rev. D",
    volume = "95",
    number = "9",
    pages = "095029",
    year = "2017"
}

@article{Marcano:2019rmk,
    author = "Marcano, Xabier and Morales, Roberto A.",
    title = "{Flavor techniques for LFV processes: Higgs decays in a general seesaw model}",
    eprint = "1909.05888",
    archivePrefix = "arXiv",
    primaryClass = "hep-ph",
    reportNumber = "IFT-UAM/CSIC-19-110, FTUAM-19-17, LPT-Orsay-19-30",
    doi = "10.3389/fphy.2019.00228",
    journal = "Front. in Phys.",
    volume = "7",
    pages = "228",
    year = "2020"
}

@article{PILAFTSIS199268,
title = {Lepton flavour nonconservation in H0 decays},
journal = {Physics Letters B},
volume = {285},
number = {1},
pages = {68-74},
year = {1992},
issn = {0370-2693},
doi = {https://doi.org/10.1016/0370-2693(92)91301-O},
url = {https://www.sciencedirect.com/science/article/pii/037026939291301O},
author = {Apostolos Pilaftsis},
}

@article{Arroyo-Urena:2013cyf,
    author = "Arroyo-Ure{\~n}a, Marco A. and Diaz-Cruz, J. Lorenzo and D{\'\i}az, Enrique and Orduz-Ducuara, Javier A.",
    title = "{Flavor violating Higgs signals in the Texturized Two-Higgs Doublet Model (THDM-Tx)}",
    eprint = "1306.2343",
    archivePrefix = "arXiv",
    primaryClass = "hep-ph",
    doi = "10.1088/1674-1137/40/12/123103",
    journal = "Chin. Phys. C",
    volume = "40",
    number = "12",
    pages = "123103",
    year = "2016"
}

@article{Arroyo-Urena:2020mgg,
    author = "Arroyo-Ure{\~n}a, M. A. and Valencia-P{\'e}rez, T. A. and Gait{\'a}n, R. and Montes De Oca, J. H. and Fern{\'a}ndez-T{\'e}llez, A.",
    title = "{Flavor-changing decay $h\to \tau\mu$ at super hadron colliders}",
    eprint = "2002.04120",
    archivePrefix = "arXiv",
    primaryClass = "hep-ph",
    doi = "10.1007/JHEP08(2020)170",
    journal = "JHEP",
    volume = "08",
    pages = "170",
    year = "2020"
}

@article{Arroyo-Urena:2023vfh,
    author = "Arroyo-Ure{\~n}a, M. A. and D{\'\i}az-Cruz, J. Lorenzo and F{\'e}lix-Beltr{\'a}n, O. and Zeleny-Mora, M.",
    title = "{Lessons from LHC on the LFV Higgs decays h{\textrightarrow}{\ensuremath{\ell}}a{\ensuremath{\ell}}b in the two-Higgs doublet models}",
    eprint = "2308.01380",
    archivePrefix = "arXiv",
    primaryClass = "hep-ph",
    doi = "10.1142/S0217751X24500799",
    journal = "Int. J. Mod. Phys. A",
    volume = "39",
    number = "21",
    pages = "2450079",
    year = "2024"
}

@article{Crivellin:2013wna,
    author = "Crivellin, Andreas and Kokulu, Ahmet and Greub, Christoph",
    title = "{Flavor-phenomenology of two-Higgs-doublet models with generic Yukawa structure}",
    eprint = "1303.5877",
    archivePrefix = "arXiv",
    primaryClass = "hep-ph",
    doi = "10.1103/PhysRevD.87.094031",
    journal = "Phys. Rev. D",
    volume = "87",
    number = "9",
    pages = "094031",
    year = "2013"
}

@article{Arroyo-Urena:2025gcs,
    author = "Arroyo-Ure{\~n}a, Marco A. and Herrera-Chac{\'o}n, Edwin A. and Melendez-Hern{\'a}ndez, Iran and Rosado-Navarro, Sebastian",
    title = "{Lepton flavor violation in Higgs boson decays at the HL-LHC}",
    eprint = "2511.19337",
    archivePrefix = "arXiv",
    primaryClass = "hep-ph",
    doi = "10.1007/JHEP04(2026)120",
    journal = "JHEP",
    volume = "04",
    pages = "120",
    year = "2026"
}

@article{Arroyo-Urena:2018mvl,
    author = "Arroyo-Ure{\~n}a, M. A. and D{\'\i}az-Cruz, J. L. and Tavares-Velasco, G. and Bola{\~n}os, A. and Hern{\'a}ndez-Tom{\'e}, G.",
    title = "{Searching for lepton flavor violating flavon decays at hadron colliders}",
    eprint = "1801.00839",
    archivePrefix = "arXiv",
    primaryClass = "hep-ph",
    doi = "10.1103/PhysRevD.98.015008",
    journal = "Phys. Rev. D",
    volume = "98",
    number = "1",
    pages = "015008",
    year = "2018"
}

@techreport{CMS-NOTE-2011-005,
      collaboration = "TheATLAS, TheCMS, TheLHCHiggsCombinationGroup",
      title         = "{Procedure for the LHC Higgs boson search combination in Summer 2011}",
      institution   = "CERN",
      reportNumber  = "CMS-NOTE-2011-005, ATL-PHYS-PUB-2011-11",
      address       = "Geneva",
      year          = "2011",
      url           = "https://cds.cern.ch/record/1379837",
}
